\newenvironment{eq}[1]
{\[\begin{array}{#1}}{\end{array}\]}

\let\rvec=\vec        

\def\ar{{\od,\and}}

 \def\({\Bigl(}
\def\){\Bigr)}
 \def\|{\Big|}
\def\then{\Rightarrow}
 \def\o{\circ}
\def\m{\bullet}
\def\x{\times}

\def\ox{\otimes}

\def\pl{{~\oplus~}}

\def\PL{\displaystyle \bigoplus}
\def\SUM{\displaystyle \sum}
\def\PROD{\displaystyle \prod}

\def\mid{\big\bracevert}

\def\sub{\subseteq}
\def\subnoteq{\subset}
\def\sup{\supseteq}

\def\and{\wedge}

\def\od{\vee}

\def\OD{\displaystyle\bigvee}

\def\rin{{\,\in\kern-.42em\in}}

\def\Res#1{\mathop{\rm Res}_{#1}\limits}

\def\tr{{\,{\rm tr }\,}}
\def\det{\,{\rm det }\,}

\def\Ad{\,{\rm Ad}\,}

\def\deg{\,{\rm deg}\,}

\def\sx{~\rvec\x~\!}

\def\A{{\,{\rm A\kern-.55emA}}}
\def\B{{\,{\rm I\kern-.2emB}}}
\def\C{{\,{\rm I\kern-.55emC}}}
\def\E{{\,{\rm I\kern-.2emE}}}
\def\G{{\,{\rm I\kern-.55emG}}}
\def\H{{{\rm I\kern-.2emH}}}
\def\I{{\,{\rm I\kern-.2emI}}}
\def\K{{\,{\rm I\kern-.2emK}}}
\def\L{{\,{\rm I\kern-.2emL}}}
\def\M{{\,{\rm I\kern-.16emM}}}
\def\N{{\,{\rm I\kern-.16emN}}}
\def\Q{{\,{\rm I\kern-.5emQ}}}
\def\R{{{\rm I\kern-.2emR}}}
\def\S{{\,{\rm I\kern-.42emS}}}
\def\T{{\,{\rm I\kern-.37emT}}}
\def\UU{{\,{\rm I\kern-.51emU}}}
\def\Z{{\,{\rm Z\kern-.32emZ}}}

\def\p{\partial}

\def\al{\alpha}  \def\be{\beta} \def\ga{\gamma}
\def\de{\delta}  \def\ep{\epsilon}  \def\ze{\zeta}
\def\th{\theta}   \def\vth{\vartheta} 
   \def\la{\lambda}   \def\si{\sigma}
\def\De{\Delta}    \def\Om{\Omega}
\def\phi{\varphi}
 \def\Ga{\Gamma}  
\def\Si{\Sigma}    \def\La{\Lambda}

\def\mod#1{\underline{\bf mod}_{#1}}

\def\vec#1{\underline{\bf vec}_{#1}}

\def\GL{{\bf GL}}
\def\SL{{\bf SL}}
\def\U{{\bf U}}

\def\O{{\bf O}}
\def\SU{{\bf SU}}

\def\SO{{\bf SO}}

 \def\D{{\bl D}}

\def\d#1{{\check{#1}}}

\def\rstate#1{|#1\rangle}

\def\brack#1{\lbrack#1\rbrack}

\def\ro#1{{\rm #1}}
\def\bl#1{{\bf {#1}}}
\def\cl#1{{\cal #1}}
\def\ul#1{\underline{#1}}

\def\ol#1{\overline{#1}}

\def\dprod#1#2{\langle#1,#2\rangle}

\def\map{\longrightarrow}
\def\inmap{\hookrightarrow}
\def\lrmap{\leftrightarrow}

\def\mape{\longmapsto}

\documentclass[12pt]{article}
\advance\topmargin by -1.6cm

\begin{document}

\begin{titlepage}

\hfill MPI-PhT/2003-15

\vskip3cm
\centerline{\bf MATTER AS SPECTRUM}
\vskip5mm
\centerline{\bf OF SPACETIME REPRESENTATIONS}
\vskip2cm
\centerline{
Heinrich Saller\footnote{\scriptsize
saller@mppmu.mpg.de}
}
\centerline{Max-Planck-Institut f\"ur Physik}
\centerline{Werner-Heisenberg-Institut}
\centerline{M\"unchen (Germany)}
\vskip25mm

\centerline{\bf Abstract}
\vskip8mm

Bound and scattering state Schr\"odinger  functions of nonrelativistic quantum mechanics
as representation matrix elements of space and time
are embedded into residual representations of
spacetime as generalizations of Feynman propagators.
The representation invariants arise as singularities
of rational representation functions
in the complex energy and
complex momentum plane. The homogeneous space
$\GL(\C^2)/\U(2)$ with rank 2, the orientation manifold of the unitary
hypercharge-isospin group,
is taken as model of nonlinear spacetime. Its representations
are characterized by two continuous invariants whose ratio will be related to
gauge field coupling constants
as residues of the related representation functions. Invariants of product representations
define
unitary Poincar\'e group representations with masses for free particles
in  tangent Minkowski spacetime.

\end{titlepage}

\newpage

\tableofcontents

\newpage

\section{Introduction}

In Wigner's  classification\cite{WIG} of the unitary irreducible Poincar\'e group
re\-pre\-sen\-ta\-tions the particles are
 characterized by  two invariants - a
mass $m^2$ for translations and  a spin (polariziation) $J$ for
rotations. Therewith, linear spacetime and free particles originate from
one operational concept, from a group and its representations.
Why the free particles have the observed masses, spins
and charges $z$ for the additional internal operations, that
is not  explained by classifying the representations of linear spacetime.
The actual spectrum of matter $(m^2,2J,z)\in\R\x\N\x\Z$ has to be understood
by additional structures, e.g. by representation
invariants of nonlinear spacetime. A related attempt is given in this paper.

The re\-pre\-sen\-ta\-tion classes of the additive group
$\R^d$ (translations) are  its characters -
 energies
   for   time translations  $\R$
and  momenta for  position translations  $\R^3$.
The translation characters
constitute the dual group $\d \R^d$ (dual space)
and give rise to   convolution algebras  of
energy and  momentum distributions and functions.
A homogeneous spacetime manifold with
tangent Minkowski  translations $x\in \R^4$
is  re\-pre\-sen\-table by
residues\cite{S002} of Fourier transformed
ener\-gy-mo\-men\-tum $q\in\d\R^4$  distributions.
The representation characterizing invariants arise as poles in the complex
energy and complex momentum plane.
Product representations come with convoluted
ener\-gy-mo\-men\-tum distributions and functions.

In  Feynman  propagators\cite{S96}
as  tempered distributions,
the Dirac ener\-gy-mo\-men\-tum
distributions on the mass shell
$\vth(\pm q_0)\de(q^2-m^2)$ describe free particles, acted upon by
unitary representations of the Poincar\'e group,
e.g. $e^{iq_0t}{\sin|\rvec q|r\over r}$, $q_0^2-\rvec q^2=m^2$.
The principal value distribution ${1\over q^2_\ro P-m^2}$
describes also interactions, e.g.
Yukawa interactions in  $e^{iq_0t}{e^{-|Q|r}\over r}$, $q_0^2+Q^2=m^2$.
In Feynman integrals as
convolutions  of ener\-gy-mo\-men\-tum distributions
the on-shell parts with the matrix elements of
unitary spacetime translation representations
give product representation matrix elements, i.e.
products of free  states.
The causally supported parts
 with the off-shell contributions, i.e. the Yukawa interactions
 with  nonunitary  position  representations,
are not convolutable. This is the origin of the
`divergence' problem in quantum field theories with interactions.

Representations of spacetime embed time and position representations.
The compact  time representations induce\cite{WIG,MACK,FOL} compact
representations
of spacetime translations, related to free particles.
The noncompact  position  representations\footnote{\scriptsize
Some people find it surprising that unitary representations
$e^{imt}\in\cl S'(\R)$ are no
elements of a Hilbert space with its unitary product in contrast to
non-unitary representations $e^{-|mz|}\in L^2(\R)\subnoteq\cl S'(\R)$.}
as seen in Hilbert space valued Schr\"odinger functions,
e.g. $e^{-|m|r}
=\int {d^3q\over \pi^2}{|m|\over
(\rvec q^2+m^2)^2}e^{-i\rvec q\rvec x}$,
induce Lorentz compatible representations of the spacetime translation
future cone which is taken as model of nonlinear spacetime\cite{S97,S991,S011}.
The position representations  are embedded into
causally supported  contributions.
Those parts do not describe free particles, they
are used for wave functions of particles
as their  `inner structure'. The invariant
 mass for the representation of the position degree of
freedom comes in a higher order pole, e.g. ${1\over (q^2-m^2)^2}$.
The representation invariant cannot be interpreted as a mass
for a free  particle.

After the discussion of time representations (harmonic oscillator),
position representations (Schr\"odinger wave functions)
and spacetime translation representations (Feynman propagators), all in the
language of residual representations with rational complex functions,
representations of nonlinear spacetime are given
and  an attempt is made to derive particles as product representations
of spacetime.

In the following, I have included, for better readability, many familiar
explicit calculations. The special functions are used as given in
the book of N.Ja.  Vilenkin and  A.U. Klimyk\cite{VIL}.

\section
[Residual Representations of Symmetric Spaces]
{Residual Representations\\of Symmetric Spaces}

Representation matrix elements
of a real finite dimensional symmetric space
$G/H$ with a Lie subgroup $H\sub G$
are complex functions thereon
\begin{eq}{rl}
g:(G/H)_{\rm repr}\map \C,&x\mape g( x)\cr
k\in G:&g_k(x)=g(k\m x)\cr
\end{eq}The symmetric space is assumed to have a {\it canonical
 pa\-ra\-me\-tri\-zation} by
an orbit in a  real vector space $V$
\begin{eq}{l}
 x\in G\m x_0\cong G/H, ~~G\m x_0\sub V\cong \R^d
\end{eq}e.g., a group by
 its Lie algebra $G=\exp L$ like $\SU(2)\cong\{e^{i\rvec\si\rvec x}\mid
 \rvec x\in\R^3\}$
 or the symmetric space
 $\SO_0(1,3)/\SO(3)\cong\{x\in\R^4\mid x^2={1\over m^2}\ne0\}$
 by the vectors of a timelike orbit (hyperboloid).

With the dual group
$q\in V^T\cong\d \R^d$ the  re\-pre\-sen\-ta\-tion classes for $G/H$
are characterizable by $G$-invariants $\{I_1,\dots ,I_r\}$,
 rational   for a compact and
rational or continuous  for
a noncompact Cartan subgroup.
The invariants are given
by $q$-polynomials
and can be built by linear invariants $q=m$
for an abelian group
 and by quadratic invariants $q^2=\pm m^2$ for  selfdual  groups.
All energy and momentum invariants will be written in mass units.

Using an appropriate generalized function
$\tilde g$ on the dual group $V^T\cong\d\R^d$
the irreducible $\U(1)$-re\-pre\-sen\-ta\-tions
$e^{i\dprod qx}$ of the tangent space
Fourier transform  $\tilde g$
 to a matrix element $g$  of the symmetric space re\-pre\-sen\-ta\-tion
\begin{eq}{rll}
(G/H)_{\rm repr}&\map \C,& x\mape g(x)=\int d^d q~ \tilde g(q)e^{iqx}
\end{eq}The functions
$\tilde g$ come as  quotient of two polynomials
where the invariant zeros of the denominator polynomial $P(q)$
characterize an
irreducible re\-pre\-sen\-ta\-tion via a Cartan subgroup representation
\begin{eq}{l}
\tilde g(q)\cong {Q(q)\over P(q)}\sim
\left\{\begin{array}{lll}
{1\over q -\mu},&\mu\in\R\pl i\R,&\hbox{linear}\cr
{(q)^j\over (q^2-m^2)^n}&m\in\R,&\hbox{compact}\cr
{(q)^j\over (q^2+m^2)^n},&m\in\R,&\hbox{noncompact}\cr\end{array}\right.
\end{eq}$ g$  is called a {\it residual re\-pre\-sen\-ta\-tion\cite{S002} of $G/H$},
the complex rational function $q\mape \tilde g(q)$
a residual representation function. Many examples are given  below.

Residual representations for the tangent space
$\log G/H=\log G/\log H$ of
a symmetric space $G/H$ will be formulated below.

A re\-pre\-sen\-ta\-tion of a symmetric space $G/H$
contains
re\-pre\-sen\-ta\-tions of subspaces $K$, e.g. of subgroups
$\SO(2)\subnoteq\SO(3)$
or  $\SO_0(1,1)\subnoteq\SO_0(1,3)/\SO(3)$.
A residual $G/H$-re\-pre\-sen\-ta\-tion
with canonical
tangent space  parameters $x=(x_K,x_\perp)$
has a {\it projection}  to a
residual $K$-re\-pre\-sen\-ta\-tion
by integration $\int d^{d-s} x_\perp$
over the complementary  space\footnote{\scriptsize
$\log G$ denotes the Lie algebra of the Lie group $G$.}
 ${\log G/H\over\log K}\cong\R^{d-s}$ - in both
examples above the 2-sphere ${\SO_0(1,3)/\SO(3)\over\SO_0(1,1)}\cong\Om^2\cong
\SO(3)/\SO(2)$
\begin{eq}{l}
K\map\C,~~ x_K\mape  g( x_K,0)

=\int {d^{d-s} x_\perp\over(2\pi)^{d-s}} g(x)
=\int d^s q_K ~\tilde g(q_K,0) e^{i  q_Kx_K }
\end{eq}The integration picks up the Fourier components
for trivial tangent space forms (momenta) $q_\perp=0$ of
${\log G/H\over\log K}$.
More explicit examples are given below.

The method of residual representations tries to translate
the relevant represnetation structures - invariants, Lie algebras, product
representations etc. - into the language of rational complex functions
$\C\ni q\mape  {Q(q)\over P(q)}\in \ol\C$ with its poles and its residues.

\section[Residual Representations of the Reals]
{Residual Representations of the Reals}

The simplest case of residual representations
is realized  by  time re\-pre\-sen\-ta\-tions
with  energy  functions (distributions)
and 1-di\-men\-sio\-nal position re\-pre\-sen\-ta\-tions
with  momentum functions (distributions) in the real 1-di\-men\-sio\-nal
compact group
$\U(1)=\exp i\R$ and noncompact group\footnote{\scriptsize
With two symbols
 for the isomorphic Lie groups $\R\cong\D(1)$,
 both a multiplicative and additive notation can be used. Therewith,
 one has different notations for the Lie group $\D(1)$ and its Lie
 algebra $\R=\log\D(1)$.
} $\D(1)=\exp \R$ with
their selfdual doublings $\SO(2)$ and $\SO_0(1,1)$ resp.

\subsection{Nondecomposable Representations of $\R$}

The nondecomposable
representations\cite{BOE,S89} of the noncompact totally ordered group $\R$
are the product of  an irreducible  factor and a nil-factor
\begin{eq}{rl}
\R\ni x\mape e^{i(\mu+\cl N_N )x}&\in\GL(\C^{1+N}), ~~N=0,1,2,\dots \cr
e^{i\mu x}&\in\GL(\C),~~e^{i\cl N_N x}\in\SL(\C^{1+N})
\end{eq}$N$ is called the {\it nildimension}.
The irreducible 1-di\-men\-sio\-nal representations $x\mape e^{i\mu x}$
with $N=0$ are  compact for $\R\map\U(1)$
with  real {\it invariant} $\mu$
or noncompact
for $\R\map\D(1)$ with  imaginary invariant.
The matrix elements of the nil-factor with nilpotent matrix
 $\cl N$ involve powers in the Lie parameter up to order $N$
\begin{eq}{l}
\hbox{e.g. }
\cl N_3={\scriptsize\pmatrix{
0&1&0&0\cr
0&0&1&0\cr
0&0&0&1\cr
0&0&0&0\cr}}\then \cl N_3^3\ne 0,~~\cl N_3^4=0 ,~~
e^{i\cl N_3x}={\scriptsize\pmatrix{
1&ix&{(ix)^2\over 2!}&{(ix)^3\over 3!}\cr
0&1&x&{(ix)^2\over 2!}\cr
0&0&1&ix\cr 0&0&0&1\cr}}
\end{eq}The representation space of a  nondecomposable
$\R$-representations can be spanned by $(1+N)$ principal vectors wherefrom
 only
one can be chosen as an  eigenvector.

The irreducible time or spacetime representations
in the quantum probability inducing group $\U(1)$ are used for particles (states) with the
eigenvalue $m\in\R$ as energy or mass.
Nondecomposable, reducible
representations come with indefinite unitary groups which cannot be
used for a probability interpretation. Therefore the
principal vectors involved - also the one eigenvector -
cannot be used to describe particles in quantum theory\cite{BRS,KO,S912,S913}.

The product of
 nondecomposable, reducible
representations can contain irreducible ones, e.g.
\begin{eq}{l}
e^{im_1x}{\scriptsize\pmatrix{
1&ix\cr
0&1\cr}}\ox
e^{im_2x}{\scriptsize\pmatrix{
1&ix\cr
0&1\cr}}\cong
e^{i(m_1+m_2)x}
{\scriptsize\left(\begin{array}{c|ccc}
1&0&0&0\cr\hline
0&1&ix&{(ix)^2\over 2}\cr
0&0&1&ix\cr
0&0&0&1\cr\end{array}\right)}
\end{eq}

The order structure of the reals defines the additive cones
(monoids) $\R_\ar$
and the bicone (bimonoid) $\R_\od\uplus\R_\and\cong\R_\od\x\I(2)$ which is set-isomorphic
to the group $\R$.
The bicone
representations come with a
trivial or faithful representation of the sign $\ep(x)={x\over|x|}\in
\I(2)=\{\pm1\}$, the cone representation matrix elements use
Heaviside's step function $\vth(\pm x)={1\pm\ep(x)\over2}$.

Therewith the $\R$-representation matrix elements are complex
linear combinations of the $\R$-functions
\begin{eq}{l}
\vth(\pm x)x^Ne^{i\mu x}=({\p\over\p i \mu})^N \vth(\pm x)e^{i\mu x}
,~~N=0, 1,\dots,~~\mu\in\R\pl i\R
\end{eq}The nilpotent powers arise by derivations with respect to the invariant.

\subsection
{Rational Complex Representation Functions}

An irreducible $\U(1)$ re\-pre\-sen\-ta\-tion of the group $\R$ - formulated
in this subsection in an application  for time $t\in\R$ and energy  - can be written
as  a {\it residue of a rational complex energy function}
or, equivalently, with  a {\it Dirac  distribution} supported by
the invariant energy $m\in\R$
\begin{eq}{rl}
\R\ni t&\mape  e^{ imt}
=\oint ~{dq\over2 i\pi}{1\over q-m}e^{ iqt}=\int dq~\de(q-m)e^{ iqt}
\in\U(1)\cr
\R\ni 0&\mape  1
=\oint ~{dq\over2 i\pi}{1\over q-m}
\cr
\end{eq}This
gives the prototype of a   residual re\-pre\-sen\-ta\-tion.
The integral $\oint$ circles the singularity
in the  mathematically positive direction.

For the group $\D(1)\cong \R$, where the dimension coincides with the rank and
where the eigenvalues $q$ are the group invariants $m$, the transition
to the residual form is a trivial transcription to the singularity $q=m$.
This will be different for groups with dimension strictly larger than rank, e.g. for
the space rotations $\SO(3)$, having dimension 3 and rank 1,
with the invariant  a square $\rvec q^2=m^2$ of the three
possible eigenvalues $\rvec q$.

The {\it Dirac  and  principal value $\ro P$ distributions}
from $\cl S'(\d\R)$ are the real    and imaginary  part resp.
of the {\it  causal   (advanced and retarded) distributions}
\begin{eq}{rl}
[ N  |m]_\ar
&=\({d\over dm}\)^ N  [0|m]_\ar
={[N|m]_\de\pm i[N|m]_\ro P\over 2},~~ N  =0,1,\dots\cr
&\cong \pm{1\over2i\pi}{\Ga(1+ N  ) \over ( q \mp io-m )^{1+ N  }}
={1\over 2}[ \de^{( N  )}(m- q )
\pm{1\over i\pi}{\Ga(1+ N  )  \over ( q_\ro P-m)^{1+ N  }}]\cr
\end{eq}

In the  Fourier transformations to $\cl S'(\R)$
the real-imaginary  decomposition goes with  the order function
decomposition $\vth(\pm t)={1\pm\ep(t)\over2}$ leading to representation matrix
elements
of future $\R_\od$ and past $\R_\and$,  of bicone and  group
\begin{eq}{rrr}
\R_\ar\ni \vth(\pm t)t\mape&
\pm
\int {dq\over 2i\pi} {\Ga(1+ N  )  \over ( q \mp io -m)^{1+ N  }}e^{iqt }
=&\vth(\pm t)(it)^ N   e^{imt}\cr
\R_\od\uplus\R_\and\ni t\mape&
\int {dq \over i\pi}~{\Ga(1+ N  )
\over ( q_{\ro P}-m)^{1+ N  }}e^{iqt }
=&\ep(t)(it)^ N  e^{imt}\cr
\R\ni t\mape&
\oint {dq \over 2i\pi}~{\Ga(1+ N  )
\over ( q-m)^{1+ N  }}e^{iqt }\hskip4mm&\cr
&=\int dq ~
\de^{( N  )}(m- q )e^{iqt }
=&(it)^ N  e^{imt}\cr
\end{eq}

All those distributions originate  from
the   representation functions in the closed complex plane
(Riemannian sphere) $\ol\C=\C\cup\infty$ with one  pole
\begin{eq}{l}
\ol\C\ni q\mape  {1
\over ( q-m)^{1+ N  }}\in\ol\C
\end{eq}The position $q=m$ and the order
$1+N$ of the singularity is related to the
continuous {\it invariant} and the
{\it dimensionality} of time re\-pre\-sen\-ta\-tion.
A trivial nildimension $N$
belongs to a {\it simple pole}
 ${1\over  q -m}$.
A possibly nontrivial  $ t^ N  $-dependence, $ N  \ge1$, is expressed
by the {\it multipoles}
${1  \over ( q -m)^{1+ N  }}$.
The {\it pole  normalization}
for the representation is given by  the residue
at the invariant
\begin{eq}{l}
\Res {m}
{\SUM_{n=0}^N}
{a_{-1-n}
\over ( q-m)^{1+ n  }}
={\SUM_{n=0}^N}
\oint{dq\over 2i\pi}{a_{-1-n}
\over ( q-m)^{1+ n  }}=a_{-1}
\end{eq}The complex functions for $a_{-1}=1$ are appropriately normalized
for the representation of the neutral  group  element.
The Fourier transforms with combinations
of different contour directions  around the pole
represent via $\vth(\pm t)$ and $\ep(t)$ the causal structure of the reals.

The product $\m$  of nondecomposable time re\-pre\-sen\-ta\-tion matrix elements
comes with the convolution $*$ of the energy
 distributions reflecting the order and the real-imaginary structure
\begin{eq}{cl}
{\scriptsize

\begin{array}{|c||c|c||c|c|}\hline
\m&\vth(t)&\vth(-t)&1&-i\ep(t)\cr\hline\hline
\vth(t)&\vth(t)&0&\vth(t)&-i\vth(t)\cr\hline
\vth(-t)&&\vth(-t)&\vth(-t)&i\vth(-t)\cr\hline\hline
1&&&1&-i\ep(t)\cr\hline
-i\ep(t)&&&&-1\cr\hline
\end{array} }&\then\cr\cr

{\scriptsize
\begin{array}{|c||c|c||c|c|}\hline
*&[N_1|m_1]_\od&[N_1|m_1]_\and&[N_1|m_1]_\de&[N_1|m_1]_\ro P\cr\hline\hline
[N_2|m_2]_\od&[N_+|m_+]_\od&0&[N_+|m_+]_\od&-i[N_+|m_+]_\od\cr\hline
[N_2|m_2]_\and&&[N_+|m_+]_\and&[N_+|m_+]_\and&i[N_+|m_+]_\and\cr\hline\hline
[N_2|m_2]_\de&&&[N_+|m_+]_\de&[N_+|m_+]_\ro P\cr\hline
[N_2|m_2]_\ro P&&&&-[N_+|m_+]_\de\cr\hline

\end{array}}&
\hbox{  with }\begin{array}{rl}
N_+&= N_1+ N_2\cr
 m_+&=m_1+m_2\cr\end{array}\cr

\end{eq}All these distributions span a unital algebra with conjugation
with the Dirac distributions
a
unital subalgebra. The causal distributions
for the representations of the cones
$\R_\ar$ constitute nonunital subalgebras which annihilate each other.
The principal value distributions are a vector subspace with
the convolutive action of the Dirac distribution subalgebra
the Dirac distributions for the group $\R$-representations
a unital
convolution algebra.

\subsection
{Compact Invariants}

Poles at a  squared re\-pre\-sen\-ta\-tion invariant
$q^2=m^2$ (compact invariant) can be combined from
linear poles at  $q=\pm |m|$, the invariants for the dual irreducible
subrepresentations involved,
formulated in this subsection for time and energy.

In addition to the {\it causal (advanced and retarded) energy distributions}
$[m^2]_\ar$ there are the   {\it (anti-)Feynman  energy-distributions}
 $[m^2]_\pm$
(different normalization factor ${1\over2}$)
 \begin{eq}{llcrl}
  [m^2]_\ar&={[m^2]_\ep\pm i[m^2]_\ro P\over2}
&\cong& \pm{1 \over 2i\pi}
{|m|\over (q\mp io)^2 -m^2}
&=\pm{1\over 4i\pi}({1\over q\mp io-|m|}-{1\over q\mp io+|m|})
\cr
[m^2]_\pm&=[m^2]_\de\pm i[m^2]_\ro P
 &\cong&\pm{1\over i\pi}
{|m|\over q^2\mp io -m^2}&
=\pm{1\over 2i\pi}({1\over q\mp io-|m|}-{1\over q\pm io+|m|})
\end{eq}The  principal value distribution
as imaginary part is combined with
the (anti-) symmetric   Dirac distributions as real part
\begin{eq}{l}
\left.\begin{array}{lcr}
[m^2]_\ep&\cong&|m|\ep(q)\de(q^2 -m^2)\cr
[m^2]_\de&\cong &|m|\de(q^2 -m^2)\cr\end{array}
\right\}\hbox{ with }i[m^2]_\ro P\cong {1 \over i\pi}
{|m|\over q^2_\ro P -m^2}\cr

\end{eq}There arise the  Dirac distributions
 with positive and negative energy support
\begin{eq}{rl}
{\scriptsize\pmatrix{1\cr\ep(q)\cr}}\de(q^2-m^2)
&=\de_\od(q^2-m^2)
\pm \de_\and(q^2-m^2)\cr
\hbox{with }\de_\ar(q^2-m^2)&=\vth(\pm q)\de(q^2-m^2)={1\over 2|m|}\de(q\mp|m|)
\end{eq}The Fourier transforms together with those of
\begin{eq}{l}
\pm{1 \over 2i\pi}
{q\over (q\mp io)^2 -m^2},~~
\pm{1 \over i\pi}
{q\over q^2\mp io -m^2}
\hbox{ etc.}
\end{eq}are
re\-pre\-sen\-ta\-tions  of the cones
and the group  with $\I(2)\x \SO(2)$ matrix elements
\begin{eq}{rrrr}
\hbox{causal:}&\R_\ar\ni\vth(\pm t)t \mape&
 \pm \int {d q\over 2i\pi}~{{\scriptsize\pmatrix{  |m| \cr
 q\cr }}\over (q\mp io)^2-m^2}
 e^{iqt}=&
 \vth(\pm t){\scriptsize\pmatrix{ i\sin |m|t \cr \cos mt\cr}}\cr
\hbox{Feynman:}&\R_\od\uplus \R_\and\ni t\mape&
 \pm\int {d q\over i\pi} ~{{\scriptsize\pmatrix{ |m|\cr q\cr }}
\over q^2\mp io-m^2} e^{iqt}=&{\scriptsize\pmatrix{1\cr\pm\ep(t)\cr }}
 e^{\pm i |mt|}\cr
\hbox{bicone:}&\R_\od\uplus\R_\and\ni t\mape&   \int {d q\over i\pi}~
{ {\scriptsize\pmatrix{|m|\cr  q\cr }}\over q_\ro P^2-m^2}e^{iqt}
=&\ep(t) {\scriptsize\pmatrix{i\sin |m|t \cr\cos mt\cr }}\cr
\hbox{group:}&\R\ni t\mape&   \int d q~ {\scriptsize\pmatrix{|m|\cr q\cr  }}
 \ep(q)\de(q^2-m^2)e^{iqt}=&
 {\scriptsize\pmatrix{i\sin |m|t \cr\cos mt\cr }}\cr
\hbox{group:}&\R\ni t\mape& \int d q~
{\scriptsize\pmatrix{|m|\cr q\cr }}\de(q^2-m^2)e^{iqt}=&
{\scriptsize\pmatrix{\cos mt\cr i\sin |m|t \cr  }}\cr
\end{eq}

By derivation with respect to the invariant there arise  distributions
with nontrivial nildimensions
$ {1\over (q^2-m^2)^{1+N}}$.

The convolution properties can be read off the time function multiplication.
 The   Feynman energy-distributions
combine real-imaginary
and order properties of time $t$ and energies $m^2$ as follows
\begin{eq}{c}
{\scriptsize\begin{array}{|c||c|c|}\hline
*&[m^2_1]_+&[m^2_1]_-
\cr\hline\hline
[m^2_2]_+&[m^2_+]_+&
{\scriptsize \begin{array}{r}
\vth(m_1^2-m_2^2) [m_-^2]_-\cr
+\vth(m_2^2-m_1^2)[m_-^2]_+\end{array}}
\cr\hline
[m^2_2]_-&
&[m^2_+]_-\cr
\hline
\end{array}}\hbox{ with } m_\pm=|m_1|\pm |m_2|
\end{eq}The Feynman distributions $[m^2]_\pm$
for the  bicone representations
form unital subalgebras.
In contrast to the advanced and retarded distributions $[m^2]_\ar$
they do not annihilate each other.

\subsection
{Noncompact Invariants}

The  functions
with imaginary poles from  a negative
squared re\-pre\-sen\-ta\-tion invariant
$q^2=-m^2$ (noncompact invariant)
\begin{eq}{l}
[-m^2]\cong {1\over \pi}{|m|\over
q^2+m^2}
\end{eq}give, by their Fourier transforms,
bicone  re\-pre\-sen\-ta\-tions
with noncompact $\D(1)$
matrix elements, valued in
the convolution algebra\footnote{\scriptsize
The convolution algebra $L^1(G)$ of a Lie group
coincides, for a  finite group,  with the  group algebra $\C^G$.}
 $L^1(\R)$ and the Hilbert space $L^2(\R)$  - formulated in this subsection for 1-di\-men\-sio\-nal
position $z\in\R$ and momentum $q\in\d\R$
\begin{eq}{l}
\R_\od\x\I(2)\ni z\mape\left\{\begin{array}{rl}
\int {d q\over\pi}~{|m|\over q^2+m^2} e^{-iqz}&=
\oint {d q\over2i\pi}[{\vth(-z)\over q-i|m|}
-{\vth(z)\over q+i|m|}]e^{-iqz}
=e^{-|  mz|}\cr
\int {d q\over\pi}~{iq\over
q^2+m^2} e^{-iqz}
&=\ep(z) e^{-| m z|}\cr
\int {d q\over\pi}~{2m^2 iq\over
(q^2+m^2)^2} e^{-iqz}
&=|m|z e^{-| m z|}\cr
\end{array}\right.
\end{eq}The representation relevant
residues are taken at  imaginary `momenta' $q=\pm i|m|$ in the complex
momentum plane.

The momentum functions
 constitute a real unital  convolution algebra
 \begin{eq}{l}
[-m_1^2]*[-m_2^2]= [-m_+^2]\cr
\end{eq}

The residues of the   complex  representation
functions
for compact (real) and noncompact (imaginary) invariant
  $\mu\in\{\pm|m|,\pm i|m|\}$ are
\begin{eq}{l}
\mu\in\C:~
\Res{\mu}{ 2\over q^2-\mu^2}
={1\over \mu},~~
\Res{\mu}{ 2q\over q^2-\mu^2}
=1
\end{eq}Higher order pole residues are obtained by $\mu^2$-derivations.

The residual normalization for
the unit element of the group, possible
for compact and noncompact invariant,
is different from a Hilbert space normalization,
possible for a noncompact invariant
only, e.g.
$\int_{-\infty}^\infty  dz~
e^{-|mz|}={2\over|m|}$.

\section
[Residual Representations of 3-Di\-men\-sio\-nal Position\\
(Spherical Waves and Yukawa Potentials)]
{Residual Representations\\of 3-Di\-men\-sio\-nal Position\\
(Free Particles and Bound Waves)}

Position representations with compact invariants
$\rvec q^2=m^2$ (real momenta)
are used for wave functions of quantum mechanical free scattering states
(free particles) whereas those with noncompact invariants
$\rvec q^2=-m^2$  (imaginary `momenta')
arise in quantum mechanical bound waves.

The  representations of 1-di\-men\-sio\-nal position
with compact and noncompact invariants
can be embedded\footnote{\scriptsize
The embedding symbol $\inmap$ is not meant to imply a unique embedding.}
 into rotation
$\SO(3)$  compatible
representations of 3-di\-men\-sio\-nal position
with the radial position  $|z|\cong |\rvec x|= r\in\R_\od$
and the compact 2-sphere $\Om^2$ which extends the sign $\I(2)$
for the two hemispheres \begin{eq}{rl}
\R=\R_\od\x\I(2)\ni z&\inmap \rvec x\ni
\R^3\cong \R_\od\x\Om^2\cr
\I(2)\ni\ep(z)={z\over |z|}&\inmap {\rvec x\over r}\in\Om^2,~~r\ne 0\cr
\end{eq}In the Pauli representation for position translations by traceless
hermitian complex $2\x2$-matrices
\begin{eq}{l}
\rvec x=
{\scriptsize\pmatrix{x_3&x_1-ix_2\cr x_1+ix_2&-x_3\cr}}
=x_a\si_a\in\R^3
\end{eq}the polar decomposition
looks as follows
with $u\in\SU(2)$ for the 2-sphere $\Om^2\cong\SU(2)/\SO(2)$
\begin{eq}{rl}
\rvec x&=u({\rvec x\over r })\o
{\scriptsize\pmatrix{r&0\cr 0&-r\cr}}
\o u^\star({\rvec x\over r })\cr
u({\rvec x\over r })&={\scriptsize\pmatrix{
\cos{\th\over2}&-e^{-i\phi}\sin{\th\over2}\cr
e^{i\phi}\sin{\th\over2}&\cos{\th\over2}\cr}}
={1\over \sqrt{2r(r+x_3)}}{\scriptsize\pmatrix{
r+x_3&-x_1+ix_2\cr x_1+ix_2&r+x_3\cr}}\in\SU(2)
\end{eq}

The Fourier transformations in 3-di\-men\-sio\-nal position are related to those
in one dimension by a radial derivative which produces the Kepler
factor ${1\over r}$
\begin{eq}{l}
\int {d^3 q\over 4\pi}~ \tilde\mu(\rvec q^2) e^{-i\rvec q\rvec x}
=-{d\over d r^2 }\int dq~\tilde\mu(q^2) e^{-iq r}
,~~\rvec\p={\rvec x\over r}{d\over d r}=2\rvec x {d^2\over d r^2}
\end{eq}The integral  over the
hemisphere directed momentum modulus $q_3=\ep(q_3)|\rvec q|$ goes over all reals
$\int_{-\infty}^\infty$.
Therewith a function of 1-dimensional position $\R\ni z\mape f(|z|)$
gives a function of 3-dimensional position
$\R^3\ni \rvec x\mape -{d\over dr^2}f(r)$, in the following called
2-sphere spread.

The scalar 3-di\-men\-sio\-nal  position representations,
nontrivial for $m\ne0$, use the
Fourier transforms with $\U(1)$ and $\D(1)$ matrix elements.
For simple poles there arise
spherical waves
for real momentum poles $|\rvec q|=\pm |m|$
 and Yukawa potentials
for imaginary `momentum' poles $|\rvec q|=\pm i|m|$
\begin{eq}{rlrl}
\int{d^3 q\over 2\pi^2}{1\over
\rvec q^2\mp io-m^2}e^{-i\rvec q\rvec x}&=
{e^{\pm i |m|r }\over r  },&
  \int{d^3 q\over \pi^2}{\mp i|m|\over
(\rvec q^2\mp io-m^2)^2}e^{-i\rvec q\rvec x}&=
e^{\pm i |m|r }\cr
\int {d^3q\over 2\pi^2}{1\over
\rvec q^2+m^2}e^{-i\rvec q\rvec x}
&={e^{-|m|r }\over r  }
,&
\int {d^3q\over \pi^2}{|m|\over
(\rvec q^2+m^2)^2}e^{-i\rvec q\rvec x}
&=e^{-|m|r } \cr
\end{eq}which are the 2-sphere spreads
of the representations of 1-dimensional position $\R$
\begin{eq}{l}
{e^{-\mu r}\over r}=-{2\over\mu}{d\over dr^2}e^{-\mu r}
,~~\mu\in\{\mp i(|m|\pm io),|m|\}
\end{eq}Position derivations
produce momentum polynomials in the numerator
for nontrivial  2-sphere representations
\begin{eq}{l}
\int {d^3q\over 2\pi^2}
{i\rvec q~\over \rvec q^2+\mu^2}e^{-i\rvec q\rvec x}=
-\rvec\p{e^{-\mu r}\over r}={\rvec x\over r }
{ 1+\mu r\over r^2 }e^{- \mu r },~~
\int {d^3q\over \pi^2}
{i\rvec q~\over (\rvec q^2+\mu ^2)^2}e^{-i\rvec q\rvec x}
=-\rvec\p{e^{-\mu r}\over \mu}={\rvec x\over r }
e^{- \mu r }\cr
\end{eq}e.g. the Yukawa force for a noncompact invariant $\mu=|m|$.

Nontrivial 2-sphere properties
are represented with spherical harmonics
$({\rvec x\over r})^L=\{\ro Y^{L}_{L_3}(\phi,\th)\mid L_3=-L,\dots,L\}$,
e.g. $\ro Y^2(\phi,\th)\cong
({\rvec x\over r})^2={\rvec x\ox\rvec x\over r^2}-{1\over3}\bl 1_3$.
To  avoid the $r=0$ ambiguity
they have to be multiplied
with appropriate radial powers leading to
the harmonic polynomials
\begin{eq}{l}
(\rvec x)_{L_3}^L=
r^L\ro Y^{L}_{L_3}(\phi,\th),~~\left\{\begin{array}{ll}
\rvec{\p}^2\ro Y^L_{L_3}(\phi,\th)&={L(1+L)\over r^2}\ro Y^L_{L_3}(\phi,\th)\cr
\rvec{\p}^2(\rvec x)_{L_3}^L&=0\cr\end{array}\right.
\end{eq}The harmonic polynomials have trivial translation properties.

The scalar  contributions in position representations  come with Bessel functions
of half-integer order -
 the hyperbolic Macdonald  functions $k_L$ for noncompact invariants
 and the spherical Hankel $h_L^\pm$,
Neumann $n_L$ and Bessel $j_L$ functions for compact invariants.
They have
angular momentum $L$-independent
large distance  behavior
$(k_L(R),h_L^\pm(R))\stackrel{R\to \infty}\longrightarrow({e^{-R}\over R},{e^{\pm iR}\over R})$ and
$L$-dependent  small distance  behavior
\begin{eq}{rcrll}
k_L(R)&=&(-R)^L({1\over R}{\p\over\p R})^L{e^{-R}\over
R}
&={1\over R^{1+L}}{e^{-R}\over 2^L}{\SUM_{n=0}^L}{(2L-n)!
\over (L-n)!}{(2R)^n\over n!}\cr
&=&(\pm i)^{1+L}h^\pm(\pm iR)
&={e^{- R}\over  R},
{1+ R\over R^2}e^{- R},\dots&\stackrel{R\to0}\longrightarrow {1\over R^{1+L}}\cr
h_L^\pm(R)
&=&n_L(R)\pm i j_L(R)&={e^{\pm iR}\over R},{1\mp iR\over R^2}e^{\pm iR},\dots\cr
&&
 n_{L}( R)
 &={\cos R\over  R},{\cos R+ R\sin R\over  R^2},\dots
&\stackrel{R\to0}\longrightarrow{1\over  R^{1+L}}\cr
&&j_{L}( R)
&={\sin R\over  R},
{\sin R- R\cos R\over  R^2},\dots&\stackrel{R\to0}\longrightarrow R^L\cr

\end{eq}

To obtain residual representations,
which are defined for $r=0$, i.e. without ambiguity or even singularity,
 the momentum degree of the
numerator ${1\over(\rvec q^2)^{N(L)}}$ and the degree of the nominator polynomial
$(\rvec q)^{L}$ have to leave a nonnegative nildimension $N$
for spin $J={L\over2}$ representations.

Therewith one obtains for the position $\R^3$-representation
matrix elements
the  Dirac momentum distributions with compact invariant for spin $J$ and nildimension $N$
\begin{eq}{l}
(\rvec q)^{2J}
\de^{(N)}(m^2-\rvec q^2)
\hbox{ for }\left\{
\begin{array}{rl}
J&=0,{1\over2},1,\dots\cr
N&=0,1,2,\dots\cr
\end{array}\right.\cr
\end{eq}The Fourier
transformed Dirac momentum distributions
starting from
the simple compact representations
\begin{eq}{rrlcr}
\R^3\ni\rvec x\mape&
 \int{d^3q\over 2\pi}~
\de(\rvec q^2-m^2)e^{-i\rvec q\rvec x}&={\sin |m|r \over r}
&=&|m|j_0(|m|r)\cr
\R^3\ni\rvec x\mape &\int{d^3q\over2 \pi}~i\rvec q~
\de(\rvec q^2-m^2)e^{-i\rvec q\rvec x}&
={\rvec x\over r}~{\sin |m|r -|m|r\cos mr\over r^2}
&=&|m|^3\rvec x~{j_1(|m|r)\over|m| r}\cr
\end{eq}describe free states (free particles). They
involve spherical Bessel functions multiplied with
appropriate radial powers to yield a regular
$r\to0$ behavior
\begin{eq}{l}
\sqrt{2\over\pi}{j_{L}( R)\over R^L}=
{\cl J_{{1+2L\over2}}(R)\over
R^{1+2L\over2}}=
{\SUM_{n=0}^\infty}{ (-{R^2\over4})^n\over \Ga({3\over2}+L+n)n!}

\end{eq}

In the   {\it dipoles} with compact invariants
\begin{eq}{l}
\int{d^3q\over \pi}~
{|m|\pm\rvec q\over (\rvec q^2\mp io-m^2)^2}e^{-i\rvec q\rvec x}
=(\bl 1_2-{\rvec x\over r})e^{\mp i|m|\rvec x}
\end{eq}the Dirac distribution derivatives give
 the representations of the
compact group $\SU(2)\cong\exp(i\R)^3$ with the group functions valued in
the Hilbert space $L^2(\SU(2))\subnoteq L^1(\SU(2))$  as
subspace\footnote{\scriptsize
For compact spaces one has $L^p(T)\sub L^q(T)$ for $p\ge q$.}
 of the convolution algebra
\begin{eq}{rl}

\R^3\ni\rvec x\mape&
\int{d^3q\over \pi}~

(|m|+\rvec q)\de'(m^2-\rvec q^2)e^{-i\rvec q\rvec x}\cr
&=\cos m r-i{\rvec x\over r}\sin|m|r
=e^{-i|m|\rvec x}\in\SU(2)
\end{eq}

The representation matrix elements from the principal value  pole
 for a compact
and noncompact invariant require a sufficiently high order pole
\begin{eq}{l}
\left.\begin{array}{ll}
{(\rvec q)^{2J}\over (\rvec q^2_\ro P\mp m^2)^{2+J+N}}
&\hbox{for }J=0,1,\dots\cr
{(\rvec q)^{2J} \over(\rvec q^2_\ro P\mp m^2)^{{5\over2}+J+N}}
&\hbox{for }J={1\over2},{3\over 2},\dots\cr\end{array}\right\}
\hbox{and }N=0,1,\dots
\end{eq}They start with {\it dipoles} for the scalars,
as to be expected from the additional $\rvec q^2$-power in the Lebesque measure
$d^3q=d\Om^2\rvec q^2 d|\rvec q|$,  and with  {\it tripoles}
for the vectors
\begin{eq}{rrcr}
\R^3\ni\rvec x\mape& \int{d^3q\over \pi^2}~
{|m|\over(\rvec q_\ro P^2\mp m^2)^2}
e^{-i\rvec q\rvec x}&=&(\sin |m|r,e^{-|m|r }) \cr
& \int{d^3q\over \pi^2}~
{i\rvec q\over(\rvec q_\ro P^2\mp m^2)^2}
e^{-i\rvec q\rvec x}&=&{\rvec x\over r}~(-\cos mr,e^{-|m|r }) \cr
\R^3\ni\rvec x\mape& \int{d^3q\over \pi^2}~
{4m^2 i\rvec q\over(\rvec q_\ro P^2\mp m^2)^3}
e^{-i\rvec q\rvec x}&=&|m|\rvec x~(\sin |m|r,e^{-|m|r }) \cr
\end{eq}The dipole for the vector
is ambiguous for $r=0$.
Representation matrix elements for
nontrivial nildimension arise by derivatives $({\p\over\p|m|})^N$
producing higher order poles and additional radial powers $r^N$.

For noncompact invariant
the Fourier transforms are  valued in  the position Hilbert space
$L^2(\R^3)$ and in the convolution algebra $L^1(\R^3)$.
The scalar dipoles and the vector tripoles etc. are
position representations
by Schr\"odinger functions
\begin{eq}{rrcrl}
\rstate{1,\rvec 0}\sim & e^{-|m|r }&=&
\int{d^3q\over \pi^2}~
{|m|\over(\rvec q^2+m^2)^2}
e^{-i\rvec q\rvec x},&|m|=1 \cr
\rstate{2,\rvec 1}\sim&2|m|\rvec x~ e^{-|m|r }&=&
\int{d^3q\over \pi^2}~
{8m^2i\rvec q\over(\rvec q^2+ m^2)^3}
e^{-i\rvec q\rvec x},&|m|={1\over 2} \cr
\end{eq}They arise as  knotless waves $\rstate{k,\rvec L}$ of the
nonrelativistic hydrogen atom\cite{MESS1}
with angular momentum $\rvec L\cong(L,{L_3})$
 and principal quantum number
$k$ - the inverse of the  quantized `imaginary' momentum
$|\rvec q|=\pm i|m|$ as invariant for the position representation
\begin{eq}{l}
\rstate{k,\rvec L}\sim
(2|m|\rvec x)^L~ \ro L_{1+2L}^N(2|m|r)e^{-|m|r}\cr
E={\rvec q^2\over 2}=-{m^2\over 2},~~|m|={1\over k},~~k=1+L+N
\end{eq}The degree of the Laguerre polynomials
\begin{eq}{rl}
\ro L^N_\la(  \rho )&=( \rho ^{-\la}e^  \rho  {d\over d  \rho }
  \rho ^ {\la}e^{-  \rho })^N{\rho^N\over N!}
 ={\SUM_{n=0}^N}{\la+N \choose\la+n }{(-\rho)^n\over n!},~~\left\{
 \begin{array}{rl}
 \R&\ni \la\ne-1,-2,\dots\cr
N&=\deg\ro L_\la^N\end{array}\right.
\end{eq}is  the radial quantum number (knot number).
Nontrivial knots, i.e. nildimensions
$N=1,2,\dots$ are obtained by operating with the Laguerre polynomials
$r^Ne^{-|m|r}=\({d\over d|m|}\)^N e^{-|m|r}$,
e.g. for one knot
\begin{eq}{rrcrl}
\rstate{2,\rvec 0}\sim&(2-2|m|r) e^{-|m|r }&=&\int{d^3q\over \pi^2}~
{4|m|(\rvec q^2-m^2)\over(\rvec q^2+m^2)^3}
e^{-i\rvec q\rvec x},&|m|={1\over2} \cr
\rstate{3,\rvec 1}\sim&2|m|\rvec x~ (4-2|m|r)e^{-|m|r }&=&
\int{d^3q\over \pi^2}~
{48 m^2i\rvec q(\rvec q^2-m^2)\over(\rvec q^2+ m^2)^4}
e^{-i\rvec q\rvec x},&|m|={1\over 3} \cr
\end{eq}

The convolutions can be read off from the matrix elements, e.g.
\begin{eq}{rll}
{1\over \pi^2}~
{|m_1|\over(\rvec q^2+m_1^2)^2}
*{1\over \pi^2}~
{|m_2|\rvec q\over(\rvec q+m_2^2)^3}
={1\over \pi^2}~
{|m_+|\rvec q\over(\rvec q+m_+^2)^3},~~|m_+|=|m_1|+|m_2|
\end{eq}

The residues
 of the scalar  complex representation
 functions with the complexified
 radial degree of freedom, e.g.
\begin{eq}{l}
\R\x\Om^2\inmap\ol \C\x \Om^2\ni
\rvec q=|\rvec q|~{\rvec q\over |\rvec q|}\mape {1\over\rvec q^2-\mu^2}\in\ol\C
\end{eq}have to take into account the 2-sphere degrees of freedom
\begin{eq}{l}
\mu\in\C:~~\Res{\mu} {4\mu\over \rvec (q^2-\mu^2)^n}
=\oint_{\mu} {d^3q\over 2i\pi^2}{4\mu\over (\rvec q^2-\mu^2)^n}
=\oint_{\mu}{d q\over 2i\pi}{4\mu q^2\over (q^2-\mu^2)^n}=\left\{
\begin{array}{rl}
2\mu^2,&n=1\cr
1,&n=2\cr\end{array}\right.
\end{eq}The additional normalization factor ${1\over\pi}$ is discussed in the
next section.

The residual normalization is used for the representation of
the unit in a Cartan subgroup, e.g. $\SO(2)\subnoteq \SO(3)$
or $\SO_0(1,1)\subnoteq \SO_0(1,3)/\SO(3)$.
It is different from a quadratic form normalization,
e.g. with the invariant bilinear Killing form
of the $\SO(3)$-Lie algebra.

\section{Residual Normalizations}

For the characters $\d\R^d$ of the translations $\R^d$ with
a signature $(d-s,s)$ metric one has
 the residual normalizations\cite{GELSHIL1}
for positive and negative invariants $\mu^2$ (where the $\Ga$-functions are defined)
\begin{eq}{l}
\begin{array}{l}
\O(d-s,s)\sx\d\R^d:\cr
\mu^2,\nu\in\R\end{array}
\left\{\begin{array}{l}
  \int {d^dq\over (\pm i)^{s}\sqrt{\pi^d}}
 ~{ \Ga({d\over2}+1+\nu )\over(q^2\mp io-\mu^2)^{{d\over2}+1+\nu } }
={2\over d}\int {d^dq\over (\pm i)^{s}\sqrt{\pi^d}}
 ~{q^2~ \Ga({d\over2}+2+\nu )\over(q^2\mp io-\mu^2)^{{d\over2}+2+\nu } }\cr
= {\Ga(1+\nu)\over (\mp io-\mu^2)^{1+\nu}}
=\left\{\begin{array}{ll}
 {\Ga(1+\nu)\over (-\mu^2_\ro P)^{1+\nu}}\pm i\pi\de^{(\nu)}(\mu^2),&\nu=0,1,2,\dots\cr
 {\Ga(1+\nu)[\vth(-\mu^2)+e^{\pm i\pi(1+\nu)}\vth(\mu^2)]
 \over |\mu^2|^{1+\nu}},&\nu\ne
0,1,2,\dots\end{array}\right.\cr\end{array}\right.
\end{eq}with the relevant examples for definite
signatures (energy and momenta) and indefinite
ones for Minkowski ener\-gy-mo\-men\-ta $\d\R^{1+s}$
\begin{eq}{rl}
\left.\begin{array}{rl}
\d\R^1:&
 \pm  \int {dq~\over i \pi}
 {1\over q^2\mp io-\mu^2  }\cr

\O(3)\sx\d\R^3:&
\pm \int {d^3q\over i\pi^2}
 ~{ 1\over(\rvec q^2\mp io-\mu^2)^2 }\end{array}\right\}
&= {\vth(\mu^2)\mp i\vth(-\mu^2)\over |\mu|}\cr
\cr
\left.\begin{array}{rl}

\O(1,1)\sx\d\R^2:&
  \pm\int {d^2q\over i\pi }
 ~{1\over(q_0^2-q_3^2\mp io-\mu^2)^2 }\cr
 \O(1,3)\sx\d\R^4:&
  \mp \int {d^4q\over i\pi^2}
 ~{ 2\over(q_0^2-\rvec q^2\mp io-\mu^2)^3 }\end{array}\right\}
&= -{1\over \mu^2_\ro P}\pm i\pi\de(\mu^2)\cr
\end{eq}This shows the addional normalization factor $\pm {1\over \pi}$
if the  residues of position $\R$ are embedded
into position $\R^3\cong\R_\od\x\Om^2$.

\section
[Residual Representations of
 2-Dimensional Spacetime]
{Residual Representations \\of  2-Dimensional Spacetime}

Residual time and position representations
can be embedded into  Minkowski spacetime
representations. They employ
ener\-gy-mo\-men\-tum distributions whose Lorentz invariant singularities
determine
the embedded representations of both time and position.

2-di\-men\-sio\-nal Minkowski spacetime
in a diagonal $(2\x2)$-matrix representation
\begin{eq}{l}
 x
={\scriptsize\pmatrix{x_0+x_3&0\cr 0&x_0-x_3\cr}}
=x_0\bl 1_2+x_3\si_3\in\R^2\cong (\R_\od\uplus\R_\and)
\hskip-1mm\pl\hskip-1mm (\I(2)\x\R_\od)
\end{eq}is acted upon with the orthochronous Lorentz group
(dual dilatations)
\begin{eq}{l}
\SO_0(1,1):~~x_0\pm x_3\mape e^{\pm {\psi_3}}(x_0\pm x_3)
\end{eq}(without rotation degrees of freedom). It is
the noncompact
abelian substructure of the Lorentz group $\SO_0(1,3)$ of 4-di\-men\-sio\-nal
spacetime $\R^4$.

\subsection
{Energy-Momentum Distributions}

The   scalar  ener\-gy-mo\-men\-tum distributions -
(anti-) Feynman  and causal (advanced, retarded)
 - are distinguished by their energy $q_0$ behavior.
They are combinations of
 the (anti-)symmetric Dirac distribution with the principal
 value distribution
\begin{eq}{rrl}

\hbox{Feynman:}&\pm{1\over i\pi}{1\over q^2\mp io-m^2}&
=\de(q^2 -m^2)
\pm {1 \over i\pi}
{1\over q^2_\ro P -m^2}\cr
\hbox{causal:}&\pm{1 \over 2i\pi}{1\over (q\mp io)^2 -m^2}
&={1\over2}[\ep(q_0)\de(q^2 -m^2)
\pm{1 \over i\pi}
{1\over q^2_\ro P -m^2}]\cr
&&\hbox{with }(q\mp io)^2=(q_0\mp io)^2-q_3^2\cr
&{\scriptsize\pmatrix{1\cr\ep(q_0)\cr}}\de(q^2-m^2)
&=\de_\od(q^2-m^2)\pm \de_\and(q^2-m^2)\cr

\end{eq}Multipoles
arise by derivations with respect to the invariant $m^2$.

The Fourier transformed $d^2q=dq_0dq_3$
Dirac distribution for ener\-gy-mo\-men\-ta
\begin{eq}{rl}
\int d^2q~\de(q^2-m^2)e^{iqx}
&=-\pi\cl N_0(\sqrt{{m^2x^2\over4}})\cr
&=-\vth(x^2)\pi\cl N_0(\sqrt{{m^2x^2\over4}})
+\vth(-x^2)2\cl K_0(\sqrt{-{m^2x^2\over4}})
\cr
\end{eq}comes with the
order 0 Neumann function for real argument (timelike)
which is the  Macdonald function for imaginary argument (spacelike)
\begin{eq}{rl}
\R\ni \xi\mape
\pi\cl N_0(\xi)
&=
{\SUM_{n=0}^\infty}{ (-{\xi^2\over4})^n\over (n!)^2}
[\log{|\xi^2|\over4}+2\ga_0-2\phi(n)]=-2\cl K_0(-i\xi)\cr
\phi(0)&=0,~~\phi(n)=1+{1\over2}+\ldots +{1\over n},~~n=1,2,\dots \cr
\ga_0&=-\Ga'(1)=\lim_{n\to\infty}\brack{\phi(n)-\log n}=0.5772\ldots\cr
\end{eq}

The advanced and retarded Fourier transforms are causally supported
\begin{eq}{rl}
 \int {d^2q \over i\pi}
{1\over q^2_\ro P -m^2}
e^{iqx}
&=\ep(x_0)
 \int d^2q~\ep(q_0)\de(q^2-m^2)e^{iqx}\cr
 &= i\pi\vth(x^2)\cl E_0({m^2x^2\over4})\cr
\end{eq}They involve Bessel functions of integer order
\begin{eq}{rl}
\R\ni \xi\mape\cl E_L({\xi^2\over4})&={\cl J_L(\xi)\over({\xi\over2})^L}=
{\SUM_{n=0}^\infty}{ (-{\xi^2\over4})^n\over (L+n)!n!}
=\(-{\p\over\p{\xi^2\over 4}}\)^L\cl E_0({\xi^2\over4}),~~
L=0,1,\dots\cr
\cl E_0({\xi^2\over4})&=\cl J_0(\xi),~~
(1+L)\cl E_{1+L}({\xi^2\over4})
=\cl E_L({\xi^2\over4})+{\xi^2\over4}\cl E_{2+L}({\xi^2\over4})
\end{eq}

The Feynman propagators proper - for particles -
have first order poles - they come
with the Hankel functions $\cl H_0^\mp=\cl N_0\mp i\cl J_0$
\begin{eq}{rl}
\pm \int {d^2q \over i\pi}
{1\over q^2\mp io  -m^2}
e^{iqx}
&=-\pi\vth(x^2)\cl H_0^\mp(\sqrt{{m^2x^2\over4}})
+\vth(-x^2)2\cl K_0(\sqrt{-{m^2x^2\over4}})\cr
\end{eq}

Fourier transformed  Lorentz vectors
\begin{eq}{l}
\pm {1\over 2i\pi}
{q\over (q\mp io)^2-m^2},~~
\pm {1\over i\pi}
{q\over q^2\mp io-m^2}
\hbox{ etc. }
\end{eq}are obtained by spacetime derivation
$\p =2x{\p\over\p x^2}$, e.g.
\begin{eq}{rl}
\int {d^2q\over i\pi}~{q\over q^2_\ro P-m^2}
e^{iqx}
&=\ep(x_0)
 \int d^2q~q\ep(q_0)\de(q^2-m^2)e^{iqx}\cr

&=\pi\p~\vth(x^2)
\cl E_0({m^2x^2\over4})
= \pi{x\over2}
[\de({x^2\over4})
-\vth(x^2)m^2\cl E_1({m^2x^2\over4})]\cr
\end{eq}

\subsection{Time and Position Frames}

The partial Fourier transformations
with respect to energy and momentum
display the spacetime embedded  time and position representations
\begin{eq}{rl}
\pi g(m^2,x)&=\int d^2q~e^{iqx}~\tilde g(m^2,q)=
  \int dq_3~ e^{-i q_3x_3} g(q_0,x_0)\cr
 & =\int dq_0~ e^{iq_0x_0}[
  \vth(q_0^2-m^2) g^c(q_3,x_3)  +\vth(m^2-q_0^2) g^{nc}(iq_3,x_3)]\cr
\cr
\hbox{time:}&
\R\ni x_0\mape   g(q_0,x_0)\cr
\hbox{position:}&
\R\ni x_3\mape\left\{\begin{array}{ll}
  g(q_3,x_3)\cr
  g^{nc}(iq_3,x_3)\cr\end{array}\right.\cr
\end{eq}

{\scriptsize
\begin{eq}{c}
\begin{array}{|c||c||c|c|}\hline
&\hbox{\bf time}
&\hbox{\bf position}
&\hbox{\bf position}\cr
&\hbox{ (compact)}
&\hbox{ (compact)}
&\hbox{(noncompact)}\cr\hline
\tilde g(m^2,q)& g(q_0,x_0)& g^c(q_3,x_3)&
 g^{nc}(iq_3,x_3)\cr\hline
&q_0=\sqrt{m^2+q_3^2}&q_3=\sqrt{q_0^2-m^2}&iq_3=|Q|=\sqrt{m^2-q_0^2}\cr
\hline\hline

\hbox{\bf Lorentz scalars}&&&\cr\hline

\hfill \de(m^2-q^2)&{\cos q_0x_0\over q_0}
&{\cos q_3x_3\over q_3}  &0\cr\hline

\hfill \ep(q_0)\de(m^2-q^2)
&  i{\sin q_0x_0\over q_0}
&\ep(q_0){\cos q_3x_3 \over q_3}
&0\cr\hline

{1 \over i\pi}{1\over q^2_\ro P -m^2}
& \ep(x_0)i{\sin q_0x_0\over q_0}
& -i{\sin  q_3|x_3|\over q_3}
&  i{e^{-  |Qx_3|}\over |Q|} \cr\hline
\hline

\hbox{\bf Lorentz vectors}&&&\cr\hline

\phantom{\ep(q_0)}q\de(m^2-q^2)&{\scriptsize\pmatrix{
\hfill i\sin q_0x_0\cr
\hfill {q_3\over q_0}~\cos q_0x_0\cr}}&
{\scriptsize\pmatrix{
\hfill {q_0\over q_3}~\cos q_3x_3 \cr
\hfill i\sin q_3x_3\cr}}&0\cr\hline

q\ep(q_0)\de(m^2-q^2)&
 {\scriptsize\pmatrix{
\hfill \cos q_0x_0\cr
\hfill {q_3\over q_0}~i\sin q_0x_0\cr}}
&\ep(q_0){\scriptsize\pmatrix{
\hfill {q_0\over q_3}~\cos q_3x_3 \cr
\hfill i\sin q_3x_3\cr}}
&0\cr\hline

{1 \over i\pi}{q\over q^2_\ro P -m^2}
& \ep(x_0)
{\scriptsize\pmatrix{
\hfill \cos q_0x_0\cr
\hfill {q_3\over q_0}~i\sin q_0x_0\cr}}
& -{\scriptsize\pmatrix{
\hfill {q_0\over q_3}i\sin q_3|x_3|\cr
\hfill \ep(x_3)\cos q_3x_3
\cr}}
&{\scriptsize\pmatrix{
  \hfill i{q_0\over |Q|}\cr
\hfill  \ep(x_3)\cr}}
 e^{-  |Qx_3|}\cr\hline

\end{array}\cr\cr
\hbox{\bf Time and Position Representations for $\R^2$-Spacetime}

\end{eq}}

\noindent The higher order poles arise by derivation
\begin{eq}{l}
{\p\over\p |m|}=2|m|{\p\over\p m^2}\cong
{|m|\over q_0}{\p\over \p q_0}
\cong-{|m|\over  q_3}{\p\over \p q_3}\cong {|m|\over |Q|}{\p\over \p |Q|}\cr
\end{eq}

The   Dirac distributions
involve  time and position representations with compact invariant,
the principal value part, in addition,
also  position representations with noncompact invariant
$q_3^2=-(m^2-q_0^2)$
\begin{eq}{l}
{1\over -q^2_\ro P+m^2}
=\vth (q_0^2-m^2){1\over  q_3^2-(q_0^2-m^2)}
+\vth (m^2-q_0^2){1\over q_3^2+(m^2-q_0^2) }
\end{eq}

The {\it projection to time representations}
will be defined by the partial Fourier
transformation $\int dx_3 g(m^2,x)$ leading to
 trivial momentum $q_3=0$ (rest system), defining a
{\it time frame}. The {\it projection to  position representations}
by the partial Fourier transformation $\int dx_0 g(m^2,x)$ leads to
trivial energy $q_0=0$ and defines a {\it position  frame}
\begin{eq}{rl}
g(|m|,t)&=\int{dx_3\over 2} g(m^2,x)
=\int d^2 q
~\de(q_3)\tilde g(m^2,q)e^{iqx}\cr
g^{nc}(i|m|,z)&=\int{dx_0\over 2} g(m^2,x)
=\int d^2 q
~\de(q_0)\tilde g(m^2,q)e^{iqx}\cr
\end{eq}Time frames have real energies for free particles -
position frames have `imaginary' momenta for bound waves

{\scriptsize

\begin{eq}{c}
\begin{array}{|c||c|c|}\hline
&\hbox{\bf time frame}
&\hbox{\bf position frame}\cr
&x_0=t&x_3=z\cr
\hline
\tilde g(m^2,q)& g(|m|,t)
&
 g^{nc}(i|m|,z)\cr
\hline
&(q_0, q_3)=(|m|,0)&(q_0, q_3)=(0,i|m|)\cr
\hline\hline

\hbox{\bf Lorentz scalars}&&\cr\hline

\hfill\de(m^2-q^2)&{\cos mt\over |m|}
&0\cr\hline

\hfill\ep(q_0)\de(m^2-q^2)
& i{\sin mt\over m}
&0\cr\hline

{1 \over i\pi}{1\over q^2_\ro P -m^2}
& \ep(t)i{\sin mt\over m}
&  i{e^{-  |mz|}\over |m|} \cr\hline
\hline

\hbox{\bf Lorentz vectors}&&\cr\hline

\phantom{\ep(q_0)}q\de(m^2-q^2)&{\scriptsize\pmatrix{
i\sin |m|t\cr
0\cr}}&
0\cr\hline
q\ep(q_0)\de(m^2-q^2)&
 {\scriptsize\pmatrix{
\cos mt\cr
0\cr}}
&0\cr\hline

{1 \over i\pi}{q\over q^2_\ro P -m^2}
&
{\scriptsize\pmatrix{
\ep(t)\cos mt\cr
0\cr}}
&{\scriptsize\pmatrix{
 0\cr
 \ep(z) e^{-  |mz|}\cr}}
\cr\hline
{1 \over i\pi}{q\over (q^2_\ro P -m^2)^2}
&
{\scriptsize\pmatrix{
-{t\sin |mt|\over 2|m|}\cr
0\cr}}
&{\scriptsize\pmatrix{
 0\cr
 -z {e^{-  |mz|}\over 2|m|}\cr}}
\cr\hline

\end{array}\cr\cr
\hbox{\bf Time and Position Projection for $\R^2\cong\R\pl\R$}
\cr
\end{eq}}

\noindent In the projections
 there remain the compact time and the noncompact position representations.
 The Dirac  ener\-gy-mo\-men\-tum distributions
 embed only  time  projections
whereas the {\it principal value distributions
 embed both  time and position}  projections.
 The time representations
have nildimensions $N=0,1,\dots $ for  poles, dipoles etc.
The  position projections
arise from spacetime
distributions with causal support
$x^2\ge0$.

The   complex representation  functions for 2-di\-men\-sio\-nal spacetime, e.g.
\begin{eq}{l}
\ol \C^2\ni q\mape {q \over q^2-m^2}\in\ol \C^2
\end{eq}have energy and momentum projected residues
with real and imaginary invariants - for Lorentz scalars
\begin{eq}{lcrrl}
\Res{\pm |m|}{2\over q^2-m^2}
&=&\oint _{\pm |m|}{d^2q\over 2i\pi}\de(q_3){2\over q^2-m^2}
=&\oint_{\pm |m|} {dq\over 2i\pi}{2\over q^2-m^2}&=\pm {1\over |m|}\cr
\Res{\pm i|m|}{2\over q^2-m^2}&=&
\oint_{\pm i|m|} {d^2q\over 2i\pi}\de(q_0){2\over q^2-m^2}
=&-\oint_{\pm i|m|} {dq\over 2i\pi}{2\over q^2+m^2}&=-{1\over \pm i|m|}\cr
\end{eq}For Lorentz vectors $q=q_0\bl 1_2+q_3\si_3$ with $\tr q=2q_0$
there is a trace residue for the  energy projection
\begin{eq}{l}
\tr \Res{\pm |m|}{q\over q^2-m^2}
=\tr \oint _{\pm |m|}{d^2q\over 2i\pi}\de(q_3){q\over q^2-m^2}
=\oint_{\pm |m|} {dq\over 2i\pi}{2q\over q^2-m^2}=1\cr
\end{eq}

Starting from the projections,
the compact position representations are induced\cite{FOL} by the
compact time representations with the eigenvalues $(q_0,q_3)$
on the {\it $\SO_0(1,1)$-mass
hyperboloid} and the $\SO_0(1,1)$-measure ${dq_3\over 2\sqrt{q_3^2+m^2}}$.
The spacetime translation representation has the cardinality of
$\SO_0(1,1)$ as its overcountably infinite dimension.
The related Dirac distributions for unitary
spacetime translation representations embed free scattering waves
(free particles)
\begin{eq}{rl}
(|m|,0)\inmap (q_0,q_3)\hbox{ with }e^{i|m|t}&\inmap
e^{iq_0x_0-iq_3x_3},~~q_0^2-q_3^2=m^2\cr
\R\ni t\mape e^{i|m|t},~~\R^2\ni x&\mape\int d^2 q~\de(q^2-m^2)e^{iqx}\cr
\end{eq}The noncompact position representation matrix elements
are functions from the position  Hilbert space. They induce time representations
with the eigenvalues $(q_0,Q)$ on the {\it $\SO(2)$-mass circle}.
The spacetime embedding for the position bound waves uses
the principal value distributions
\begin{eq}{l}
(0,i|m|)\inmap (q_0,iQ)\hbox{ with }
e^{-|m z|}\inmap e^{iq_0x_0-|Qx_3|},~~q_0^2+Q^2=m^2\cr
\R\ni z\mape e^{-|mz|},~~\hbox{ causal } x^2\ge 0:
 \R^2\ni x\mape \int d^2q {1\over q_\ro P^2-m^2}
e^{iqx}\cr
\end{eq}

\subsection{Singularity Surfaces in Energy-Momenta}

For time and 1-dimensional position, the representation functions
\begin{eq}{l}
\R\inmap\ol\C\in q\mape {1\over q^2\mp m^2}\in\ol\C
\end{eq}are
singular at   points in the complex
plane $\C\cong\R^2$, at $\{\pm |m|,0)\}$ for compact and at $\{(0,\pm i|m|)\}$
for noncompact representations.
For 2-dimensional spacetime, the singularities
of
\begin{eq}{l}
\R\pl\R\inmap\ol\C\pl\ol \C=\ol\C^2\ni q\mape {1\over q^2-m^2}\in\ol\C
\end{eq}are on
a real 2-dimensional surface in the real 4-dimensional space
$\C^2\cong\R^4$
with a complex  energy and a complex momentum plane
\begin{eq}{l}
(q_0,\Ga;q_3,Q)\in\R^4:~~
\left\{\begin{array}{rl}
q_0^2-\Ga^2-q_3^2+Q^2&=m^2\cr
q_0\Ga-q_3Q&=0\cr\end{array}\right.
\end{eq}For nontrivial mass the singularity surface
can be parametrized with a positive and negative energy-like
hyperboloid
and a foward and backward  momentum-like
hyperboloid
\begin{eq}{l}
m^2>0,~~\SO_0(1,1):~\left\{\begin{array}{llll}
q_0^2-q_3^2&=m_0^2,&
(q_0,q_3)&=m_0(\cosh\psi,\sinh\psi)\cr
\Ga^2-Q^2&=-m_3^2,&(\Ga,Q)&=m_3(\sinh\psi,\cosh\psi)\cr\end{array}\right.
\end{eq}For four spacetime dimensions the momentum-like hyperboloid
has one shell only, $\ep(z)\inmap{\rvec x\over r}$.
The singularity surface contains
the circles
\begin{eq}{l}
\SO(2):~~\left\{\begin{array}{rl}
m_0^2+m_3^2&=m^2\then (m_0,m_3)=|m|(\cos\al,\sin\al)\cr
q_0^2+Q^2&=m^2\cosh^2\psi\cr
q_3^2+\Ga^2&=m^2\sinh^2\psi\end{array}\right.
\end{eq}Therewith, the singularity surface in $\C^2$
is four times a circle, embedding the
imaginary poles for noncompact $\R$-representations,
sliding along a hyperboloid which embeds the
real poles for compact $\R$-representations. It can be seen in the
$\R^3$-projection to real energies
where the energy-momentum hyperbola touches
the energy-imaginary `momentum' circle at the two points
$(\pm |m|,0;0,0)$
\begin{eq}{rcl}
\R\pl\C\ni(q_0,0;q_3,Q):
&&\{q\mid q_0^2-q_3^2=m^2,~Q=0\}\cr
&\cup& \{q\mid q_0^2+Q^2=m^2,~q_3=0\}\cr
\end{eq}and in the $\R^3$-projection
to real momenta where there is the  energy-momentum hyperbola
only
\begin{eq}{l}
\C\pl\R\ni(q_0,\Ga;q_3,0):~~\{q\mid q_0^2-q_3^2=m^2,~\Ga=0\}
\end{eq}For trival invariant  the circles shrink to points on the
hyperbola
\begin{eq}{l}
m^2=0:~~(\Ga,Q)=0\hbox{ or } (q_0,q_3)=0\then  \hbox{ trivial }\SO(2)
\end{eq}

In ${1\over q^2-m^2}$, $m^2>0$, there
is only one Lorentz invariant  for the  real 2-dimensional
{\it hyperbolic-spherical singularity surface}.
For representations of nonlinear spacetime below two invariants will be
introduced - to embed compact  representations
$e^{imt}$ and noncompact ones $e^{-|m|r}$,
each kind with an independent invariant.

\section
{Convolutions  for Spacetime}

Feynman integrals as used in perturbation theory
involve convolutions of ener\-gy-mo\-men\-tum distributions
for pointwise
products of spacetime  distributions.
In general they do not make sense
since $\cl S(\d\R^d)$ is no convolution algebra.

For ener\-gy-mo\-men\-tum convolutions the points on the hyperbolic-spherical
singularity surfaces involved are added. The
addition of compact with compact and
noncompact with noncompact invariants
embed products for
time and position representations. The characteristically
new feature is the addition of compact with noncompact invariants.

\subsection
[Convolution  of 2-Dimensional Energy-Momentum Distributions]
{Convolution  of 2-Dimensional\\Energy-Momentum Distributions}

The product of Feynman propagators for product representations
of spacetime uses the
convolution of ener\-gy-mo\-men\-tum  distributions
where  $\de ({\SUM_j}q_j-q)$
 adds up the ener\-gy-mo\-men\-ta
as spacetime translation eigenvalues
to the  eigenvalue $q$
of the product representation, e.g.
for scalar multipole Feynman propagators
\begin{eq}{l}
\pm{1\over i\pi}{\Ga(1+n_1)\over (q^2\mp io-m_1^2)^{1+n_1}}
*\cdots*
\pm{1\over i\pi}{\Ga(1+n_k)\over (q^2\mp io-m_k^2)^{1+n_k}}
 \cr
\hskip2cm=
(\pm{1\over i\pi})^k\int d^2 q_1\cdots
d^2 q_k  \de ({\SUM_{j=1}^k}q_j-q)
{\PROD_{j=1}^k}{\Ga(1+ n_j )\over (q_j^2\mp io -m_j^2)^{1+ n_j} }  \cr
\end{eq}The convoluted  Feynman distributions  have to be all of
the same type, either all advanced  $q^2-io$
or all retarded $q^2+io$.

The convolution is performed  by joining first
the invariant determining quadratic denominator
polynomials  of the energy-mo\-men\-tum distributions
\begin{eq}{rl}
 {\Ga(\nu_1)\cdots \Ga(\nu_k)\over R_1^{\nu_1}\cdots  R_k^{\nu_k}}
&=\int_0^1d\ze_1\cdots\int_0^1 d\ze_k\de(\ze_1+\dots+\ze_k-1)
{\ze_1^{\nu_1-1}\cdots\ze_k^{\nu_k-1}\Ga(\nu_1+\dots+\nu_k)
\over( R_1\ze_1+\ldots+ R_k\ze_k)^{\nu_1+\dots+\nu_k} }\cr
&\nu_j\in\R,~~\nu_j\ne0,-1,-2,\dots
\end{eq}e.g. for    two Feynman  distributions
\begin{eq}{l}
\pm{1\over i\pi}{{\scriptsize\pmatrix{1\cr q\cr}}
\Ga(1+n_1)\over (q^2\mp io-m_1^2)^{1+n_1}}
*\pm{1\over i\pi}{{\scriptsize\pmatrix{1\cr q\cr}}\Ga(1+n_2)\over (q^2\mp io-m_2^2)^{1+n_2}}\cr
\hskip10mm ={1\over i\pi}\int_0^1d\ze_{1,2}\de(\ze_1+\ze_2-1)\int {d^2p\over i\pi}
{{\scriptsize\pmatrix{1&q\ze_1\cr q\ze_2
&-p\ox p+ q\ox q\ze_1\ze_2\cr}}\ze_1^{n_1}\ze_2^{n_2} \Ga(2+ n_1+ n_2)
\over\brack{
 p^2\mp io+q^2\ze_1\ze_2  -m_1^2\ze_1 -m_2^2\ze_2 }^{2+ n_1+ n_2}   }\cr
\end{eq}For the integration the tensor
$p\ox p- q\ox q\ze_1\ze_2$ can be replaced by
$\bl 1_2{p^2\over2}- q\ox q\ze_1\ze_2$.

The convolution is the {\it $q$-dependent residue
 of the relative energy-mo\-men\-ta}
$p=q_1-q_2$
\begin{eq}{l}
\pm\int{d^2p\over i\pi}{\Ga(2+n)\over (p^2\mp io+a)^{2+n}}
=\pm\int{d^2p\over i\pi}{p^2~\Ga(3+n)\over (p^2\mp io+a)^{3+n}}
={\Ga(1+n)\over (\mp io+a)^{1+n}}\cr
\end{eq}which leads to
\begin{eq}{l}
\pm{1\over i\pi}{\Ga(1+ n_1 )\over(q^2\mp io-m_1^2)^{1+ n_1}}
*\pm{1\over i\pi}
{\Ga(1+ n_2) \over(q^2\mp io-m_2^2)^{1+ n_2}}\cr
\hskip40mm=\pm{1\over i\pi}
\int_0^1 d\ze  {\ze^{ n_1}(1-\ze)^{ n_2} \Ga(1+ n_1+ n_2)
\over[(q^2\mp io)\ze(1-\ze)-m_1^2\ze -m_2^2(1-\ze)]^{1+ n_1+ n_2} }\cr
\end{eq}Here and in the following the  convolutions  exist
only for  pole orders where the involved $\Ga$-functions are defined.
Elsewhere, there arise `divergencies'.

\subsection
[Compact and Noncompact Convolution Contributions]
{Compact and Noncompact \\Convolution Contributions}

The convolution of two  Feynman distributions
for $s$-di\-men\-sio\-nal position
$\R^s$
\begin{eq}{rl}
\pm \int{ d^{1+s}q\over i\pi}~
{1\over q^2\mp io -m^2}
e^{iqx}
= \int d^{1+s}q[1\pm\ep(q_0x_0)]\de(q^2-m^2)e^{iqx}\cr
\end{eq}gives
as real  part the difference of the  squares of  Dirac and  principal
value contributions (with $\ep(x_0)^2=1$)
whereas the imaginary part  contains the  mixed terms
\begin{eq}{l}
(\de^1\pm i\ro P^1)*(\de^2\pm i\ro P^2)=
\de^{1*2}\pm i\ro P^{1*2},\left\{\begin{array}{rl}
\de^{1*2}&=\de^1*\de^2-\ro P^1*\ro P^2\cr
\ro P^{1*2}&=\de^1*\ro P^2
+\ro P^1*\de^2\cr
\end{array}\right.\cr
\end{eq}The product of the order functions in the product of two Feynman
propagators
\begin{eq}{rccl}
[1\pm\ep(q_0x_0)][1\pm\ep(p_0x_0)]\hskip-3mm&=\hskip-3mm&
[1+\ep(q_0p_0)]\hskip-3mm&\pm[\ep(q_0)+\ep(p_0)]\ep(x_0)\cr
\hskip-3mm&=\hskip-3mm&2[\vth(q_0)\vth(p_0)+\vth(-q_0)\vth(-p_0)]
\hskip-3mm&\pm[\ep(q_0)+\ep(p_0)]\ep(x_0)\cr
\end{eq}allows the disentanglement of the
convolution
\begin{eq}{rrl}
\pm{1\over i\pi} {1\over q^2\mp io -m_1^2}
*\pm{1\over i\pi}{1\over q^2\mp io-m_2^2}
\hskip-3mm&=2\Bigl[~~&\hskip-2mm\vth(+q_0)\de(q^2-m_1^2)
*\vth(+q_0)\de(q^2-m_2^2)\cr
&+&\hskip-2mm\vth(-q_0)\de(q^2-m_1^2)
*\vth(-q_0)\de(q^2-m_2^2)\Bigr]\cr
&\hskip5mm\pm\hfill&\hskip-5mm{1\over i\pi}
\Bigl[\de(q^2-m_1^2)*{1\over q^2_\ro P-m_2^2}+
{1\over q^2_\ro P-m_1^2}*\de(q^2-m_2^2)\Bigr]
\end{eq}

The convolution  with the singularities
for nontrivial position $\R^s$ on $s$-dimensional  hyperboloids
does not lead $s$-dimensional  hyperboloids
 $\de(q^2-m_+^2)$,  but  to {\it thresholds} for  energy-mo\-men\-ta
 $q^2=(q_1+q_2)^2\ge  m_+ ^2$
 \begin{eq}{rl}
\vth(\pm q_0)\de(q^2-m_1^2)*\vth(\pm q_0)\de(q^2-m_2^2)
&\sim\vth(\pm q_0)\vth(q^2-m_+^2)\cr
\end{eq}Here, the energy is enough to produce
 two free real particles with masses $m_{1,2}$ and momentum
 $(\rvec q_1+\rvec q_2)^2\ge0$
\begin{eq}{l}
\vth(\pm q_0)\vth(q^2-m^2)
=\vth(\pm q_0)\int_0^{\rvec q^2}d \rvec p^2\de(q_0^2-\rvec p^2-m^2)\cr
\end{eq}The convolution of two step functions at masses  $m_{1,2}$
gives a step function
for the sum mass $m_+ =|m_1|+ |m_2|$.
The set with all $s+1$-dimensional
 forward (backwards)
hyperboloids $\{\{q\succeq |m|\}\mid m\in\R\}$ is an additive cone
\begin{eq}{rl}
\{q\succeq |m_1|\}
+\{q\succeq |m_2|\}
&=\{q\succeq |m_+|\}\cr
\vth(\pm q_0)\vth(q^2-m_1^2)*\vth(\pm q_0)\vth(q^2-m_2^2)
&\sim\vth(\pm q_0)\vth(q^2-m_+^2)\cr
\end{eq}The convolution of compact translation representation matrix elements
from the real part of the propagator
(free particles) gives
corresponding matrix elements
for  product representations
(product of free particles).
The positive and negative ener\-gy-mo\-men\-tum distributions
are convolution algebras, not annihilating each other
\begin{eq}{l}
\begin{array}{rlrr}
\de&=\de_\od+\de_\and,&
\de^1*\de^2=&(\de^1_\od+\de^1_\and)*(\de^2_\od+\de^2_\and)\cr
\ro P&\sim i\ep(x_0)(\de_\od-\de_\and),&
\ro P^1*\ro P^2=&
-(\de^1_\od-\de^1_\and)*(\de^2_\od-\de^2_\and)\cr
\end{array}\cr
\de^{1*2}=\de^1*\de^2-\ro P^1*\ro P^2
=2(\de_\od^1*\de_\od^2+\de_\and^1*\de_\and^2)
\sim\int_{\R^s} 2\de^{1+2}\cr
\hskip30mm\hbox{with }\de_\ar\in \cl D'(\d \R^{1+s}_\ar)\cr
\end{eq}

For time and energy, also the principal value part adds up the
invariant poles
\begin{eq}{l}
\hbox{only for time: }
\ro P^{1*2}
=\de^1*\ro P^2+\ro P^1*\de^2
\sim 2i\ep(t)(\de^1_\od*\de^2_\od-\de^1_\and*\de^2_\and)
\sim 2\ro P^{1+2}
\end{eq}The characteristic effect
of a convolution of  noncompact with compact invariant comes in
the principal value  part for $s=1,3$ position degrees of freedom
\begin{eq}{rlccc}
\de(q^2-m^2)&\sim& \vth(q^2-m^2)\cr
{1\over q_\ro P^2-m^2}
&\sim& \vth(q^2-m^2)&+&\vth(-q^2+m^2)\cr
&&\cup&&\cup\cr
& &\hbox{compact (free)}
&+&\hbox{noncompact}\cr
& &e^{imt}
&&e^{-|mz|}
\end{eq}The two  ener\-gy-mo\-men\-tum dependent zeros  of  the denominator polynomial
\begin{eq}{l}
-P(\ze)=q^2\ze(1-\ze)-m_1^2\ze -m_2^2(1-\ze)
=-q^2[\ze-\ze_1(q^2)][\ze-\ze_2(q^2)]\cr
\ze_{1,2}(q^2)={\scriptsize {q^2- m_+  m_-\pm\sqrt{\De(q^2)}\over 2q^2}   }
\hbox{ with }\left\{\begin{array}{l}
m_\pm =|m_1|\pm |m_2|\cr
\De(q^2)=(q^2- m_+ ^2)(q^2- m_-^2)\end{array}\right.
\cr\end{eq}are either both real or
complex conjugate to each other according to the sign of the
discriminant $\De(q^2)$.
Furthermore, real  zeros - in the case of $\De(q^2)\ge0$ - are in the
integration $\ze$-interval $[0,1]$
only for energy-mo\-men\-ta over the
threshold $\vth(q^2-m_+^2)$. Therewith, the convolution of
scalar propagators for 2-di\-men\-sio\-nal spacetime  reads
\begin{eq}{l}
\R^2:~\pm{1\over i\pi}{1\over q^2\mp io-m_1^2}
*\pm{1\over i\pi}
{1 \over q^2\mp io-m_2^2}=
\pm {1\over i\pi}\int_0^1  d\ze~{1\over
(q^2\mp io)\ze(1-\ze)-m_1^2\ze -m_2^2(1-\ze)}\cr\cr
\hskip8mm=
\int_0^1  d\ze\left[\de\(q^2\ze(1-\ze)-m_1^2\ze -m_2^2(1-\ze)\)
\pm {1\over i\pi}
 {1\over
q^2_\ro P\ze(1-\ze)-m_1^2\ze -m_2^2(1-\ze)}\right]\cr\cr
\hskip8mm=
{2\over\sqrt {|\De(q^2)|}}\left[\vth(q^2-m_+^2)

\mp {1\over i\pi}\vth(-\De(q^2))~
\arctan{2\sqrt{-\De(q^2)}\over  \Si(q^2)}

 \right.\cr
\hskip49.5mm\left.
\mp {1\over i\pi}~
\vth(\De(q^2))\hskip5mm\log\Big|{ \Si(q^2)-2\sqrt{\De(q^2)}
 \over  m_+^2-m_-^2} \Big|

\hskip3mm\right]\cr
\cr
\hskip8mm
 \hbox{with }
 \Si(q^2)=
(q^2 -m_+ ^2)+(q^2- m_-^2)\cr
\end{eq}The  spacetime original convolution of compact with noncompact
 invariants
 is proportional to  $\vth (-q^2+m_-^2)$ and  comes in the logarithm
\begin{eq}{rr}
\vth(-\De(q^2))=&-\vth(q^2-m_+^2)+\vth (q^2-m_-^2)\cr
\vth(\De(q^2))=&\vth(q^2-m_+^2)+\vth (m_-^2-q^2)\cr
\Big|{ \Si(q^2)-2\sqrt{\De(q^2)}
 \over  m_+^2-m_-^2}\Big|
=&\Big|{(\sqrt{m_+^2-q^2}-\sqrt{m_-^2-q^2})^2\over m_+^2-m_-^2}\Big|\hfill
\end{eq}

In the
correspondingly computed  convolution
of  energy distributions  the integral compensates the $m_-^2$-pole
from the discriminant
\begin{eq}{rl}
\R:~\pm{1\over i\pi}{|m_1|\over q^2\mp io-m_1^2}
*\pm{1\over i\pi}
{|m_2| \over q^2\mp io-m_2^2}&=
\pm {1\over i\pi}\int_0^1  d\ze~{|m_1m_2|\over
[-(q^2\mp io)\ze(1-\ze)+m_1^2\ze +m_2^2(1-\ze)]^{3\over2}}\cr
&=\pm{1\over i\pi}{|m_+|\over q^2\mp io -m_+^2}\cr
\hbox{with }{1\over P(\ze)^{3\over2}}&=
-{4\over (q^2-m_+^2)(q^2-m_-^2)}{d^2\sqrt {P(\ze)}\over d\ze^2}
\end{eq}

In the convolution of two advanced or two retarded distributions
the pole integration description has to
be changed  $q^2\mp io\to (q\mp io)^2$
everywhere
\begin{eq}{rl}
\pm \int{ d^{1+s}q\over 2i\pi}~
{1\over (q\mp io)^2 -m^2}
e^{iqx}
&= \int d^{1+s}q~\ep(q_0){1\pm\ep(x_0)\over 2}\de(q^2-m^2)e^{iqx}\cr
&= \vth(\pm x_0)\int d^{1+s}q~\ep(q_0)\de(q^2-m^2)e^{iqx}\cr
\end{eq}which antisymmetrizes  the resulting step function
above for the threshold
\begin{eq}{rl}
\pm{1\over 2i\pi}{1\over (q\mp io)^2-m_1^2}
*\pm{1\over 2i\pi}
{1 \over (q\mp io)^2-m_2^2}&=
\pm {1\over 2i\pi}\int_0^1  d\ze~{1\over
(q\mp io)^2\ze(1-\ze)-m_1^2\ze -m_2^2(1-\ze)}\cr
&={1\over 2\sqrt {|\De(q^2)}|}\left[\ep(q_0)\vth(q^2-m_+^2)
\mp {1\over i\pi}\{\dots\}\right]
\end{eq}

\subsection
{Residual Product of Representation Functions}

The convolutions of causal and Feynman ener\-gy-mo\-men\-tum distributions
can be summarized with the notation ${\stackrel{\rm R}*}$
for the different integration contours
\begin{eq}{l}
\SO_0(1,1)\sx\R^2:~~
({\stackrel{\rm R}*},q^2)=\left\{\begin{array}{ll}
(\pm{*\over i\pi},q^2\mp io),&\hbox{Feynman}\cr
(\pm{*\over 2i\pi},(q\mp io)^2),&\hbox{causal}\cr\end{array}\right.
\end{eq}with the results
\begin{eq}{l}
\R^2:~~\left\{\begin{array}{rcll}
{\Ga(1+ n_1 )\over(q^2-m_1^2)^{1+ n_1}}
&{\stackrel{\rm R}*}&{\Ga(1+ n_2) \over(q^2-m_2^2)^{1+ n_2}}&=
\int_0^1 d\ze  {\ze^{ n_1}(1-\ze)^{ n_2} \Ga(1+ n_1+ n_2)
\over[q^2\ze(1-\ze)-m_1^2\ze -m_2^2(1-\ze)]^{1+ n_1+ n_2} }\cr
{q~\Ga(1+ n_1 )\over(q^2-m_1^2)^{1+ n_1}}
&{\stackrel{\rm R}*}&
{\Ga(1+ n_2) \over (q^2-m_2^2)^{1+ n_2}}
&=\int_0^1 d\ze  {q~\ze^{ n_1}(1-\ze)^{1+ n_2} \Ga(1+ n_1+ n_2)
\over[q^2\ze(1-\ze)-m_1^2\ze -m_2^2(1-\ze)]^{1+ n_1+ n_2} }\cr
{q~\Ga(1+ n_1 )\over(q^2-m_1^2)^{1+ n_1}}
&{\stackrel{\rm R}*}&
{q~\Ga(1+ n_2) \over(q^2-m_2^2)^{1+ n_2}}
&=
\int_0^1 d\ze  { -({1\over 2}\bl1_2+q\ox q{\p\over\p q^2})
\ze^{n_1}(1-\ze)^{n_2} \Ga(n_1+ n_2)
\over[q^2\ze(1-\ze)-m_1^2\ze -m_2^2(1-\ze)]^{n_1+ n_2} }\cr\end{array}\right.
\end{eq}The convolution product
contains the normalization factor for the  relative ener\-gy-mo\-men\-tum
residue integral ${1\over 2i\pi}\oint$. Therewith,
it defines the {\it residual product} leading from
complex
 representation functions to
 functions for product representations.

The corresponding residual product
for time representations reads
\begin{eq}{l}
\R:~~\left\{\begin{array}{c}
({\stackrel{\rm R}*},q)=
(\pm{*\over 2i\pi},q\mp io)\cr
{\Ga(1+ n_1 )\over (q-m_1)^{1+ n_1}}
~{\stackrel{\rm R}*}~{\Ga(1+ n_2) \over(q-m_2)^{1+ n_2}}=
{\Ga(1+ n_1+ n_2)
\over [q-(m_1+m_2)]^{1+ n_1+ n_2} }\cr\end{array}\right.
\end{eq}The
meromorphic functions, i.e. only pole singularities,
on the closed complex plane is the field of  rational
functions. The time representation functions
$\cl P(\ol\C)$ (pole functions)
have negative degree
\begin{eq}{l}
\ol \C\ni q\mape
{P^n(q)\over P^m(q)}={a_0+a_1q+\dots+a_nq^n\over b_0+b_1q+\dots+b_mq^m}\in\ol \C,~~
a_j,b_j\in\C,~b_m\ne 0,~~n-m\le-1
\end{eq}They have a residual product with unit ${1\over q}$
adding up the invariant singularities.

The $q^2$-singularities for product representations
disappear for the
residual product of the spacetime representation pole functions.
A massless representation function ${q\over q^2}$
has  compact invariants only, i.e.
a hyperbolic singularity surface.
Its residual product
\begin{eq}{rrll}
\hbox{time }\R:&
{1\over q}
~~{\stackrel{\rm R}*}&{1\over q-m}
&={1 \over q-m}\cr
\hbox{spacetime }\R^2:&{q\over q^2}
~~{\stackrel{\rm R}*}&{q  \over(q^2-m^2)^2}
&=-({1\over 2}\bl1_2+q\ox  q {\p\over\p q^2})
\int_0^1 d\ze{1\over q^2\ze-m^2 }
\cr
&&\int_0^1d\ze{1\over q^2\ze-m^2 }
&={\log {m^2- q^2\over m^2}\over q^2}
\end{eq}gives logarithms
as integrated
 representation functions
\begin{eq}{l}
\log (1- {q^2\pm io\over m^2})
=\vth(q^2-m^2)[\pm i\pi+\log ({q^2\over m^2}-1)]
+\vth(-q^2+m^2)\log (1-{q^2\over m^2})\cr
\end{eq}The logarithm of a quotient is typical for
a finite integration\cite{BESO}, e.g. for a function
holomorphic on the integration curve
\begin{eq}{l}
\int_a^bdz~f(z)={\sum\Res{}}[f(z)\log{z-b\over z-a}],~~
\int_0^\infty dz~f(z)=-{\sum\Res{}}[f(z)\log z]
\end{eq}with the {\it sum of all residues} in the closed complex plane,
cut along the integration curve.
For 2-di\-men\-sio\-nal spacetime $\R^2\inmap\C^2$ the formulation
with the sum of the residues  looks as
follows
\begin{eq}{rl}
-\int_0^1
d\ze{1\over q^2\ze-m^2 }&=
{M^2({m^2\over q^2})\over q^2}\cr
M^2({m^2\over q^2})
=-\int_{0}^{1}
{d\ze\over \ze-{m^2\over q^2}}&=-\sum\Res{}[{1\over \ze-{m^2\over q^2}}
\log{\ze-1\over \ze}]=-\log (1- {q^2\over m^2})
\end{eq}

\section
[Residual Representations of
 4-Dimensional Spacetime]
{Residual Representations \\of  4-Dimensional Spacetime}

4-di\-men\-sio\-nal Minkowski spacetime
and its Lorentz group has
 - with 3-di\-men\-sio\-nal position
 translations $\R^3$ - additional rotation degrees of freedom
from the 2-sphere $\Om^2$.
Spacetime is used in the Cartan
representation with
hermitian  complex $(2\x2)$-matrices where
the trace is the time projection
\begin{eq}{rl}
\R\pl [\R_\od\x\I(2)]\cong\R^2&\inmap\R^4\cong\R\pl [\R_\od\x\Om^2]\cr
{\scriptsize\pmatrix{x_0+x_3&0\cr0&x_0- x_3\cr}}
&\inmap
{\scriptsize\pmatrix{x_0+x_3&x_1-ix_2\cr
x_1+ix_2&x_0-x_3\cr}}
=u({\rvec x\over r}){\scriptsize\pmatrix{x_0+r&0\cr
0&x_0-r\cr}} u^\star({\rvec x\over r})\cr
\SO_0(1,1)&\inmap \SO_0(1,3)\cong\SO_0(1,1)\x \Om^2\x\SO(2)\x\Om^2\cr
\end{eq}It   requires rotation representations
 which
 will lead, in comparison to 2-di\-men\-sio\-nal spacetime,  to  a change in the pole orders for residual
representations.

\subsection
{Feynman Distributions}

In the   Feynman and causal
ener\-gy-mo\-men\-tum distributions
\begin{eq}{rrl}
\hbox{Feynman:}&\mp{1\over i\pi^2}{\Ga(2+n)\over (q^2\mp io-m^2)^{2+n}}
&=-{1\over\pi}\de^{(1+n)}(m^2-q^2)
\mp {1\over i\pi^2}{\Ga(2+n)\over (q_{\ro P}^2-m^2)^{2+n}}\cr
&\hbox{for}&n=-1,0,1,\dots\cr
\hbox{causal:}&\mp{1\over 2i\pi^2}{\Ga(2+n)\over ((q\mp io)^2-m^2)^{2+n}}
&=-{1\over2\pi}\ep(q_0)\de^{(1+n)}(m^2-q^2)
\mp {1\over 2i\pi^2}{\Ga(2+n)\over (q_{\ro P}^2-m^2)^{2+n}}\cr
&\hbox{with}&(q\mp io)^2=(q_0\mp io)^2-\rvec q^2\cr
\end{eq}there is an additional residual normalization factor
$-{1\over\pi}$ for the 2-sphere.

The Fourier transformations
$d^4 q=dq_0 d\Om^2\rvec q^2 d|\rvec q|$ in 4-di\-men\-sio\-nal spacetime
are obtainable from the 2-di\-men\-sio\-nal case by an  invariant derivation
(2-sphere spread)
\begin{eq}{rl}
\int {d^4 q\over4\pi}~
{\scriptsize\pmatrix{1\cr\ep(q_0)\vth(q^2)\cr}}
\tilde\mu(q^2)e^{iqx}
&\hskip-3mm =-{\p\over \p  r^2}\int dq_0 d q_3~
{\scriptsize\pmatrix{1\cr\ep(q_0)\vth(q_0^2-q_3^2)\cr}}
\tilde\mu(q_0^2-q_3^2)e^{iq_0x_0-  i q_3r}\cr
&\hskip-3mm ={\p\over \p  x^2}\int d^2q~
{\scriptsize\pmatrix{1\cr\ep(q_0)\vth(q^2)\cr}}
\tilde\mu(q^2)e^{iqx}\Big|_{x=(x_0,r)}\cr
\end{eq}

One obtains as   Fourier transformation
 of the Dirac distribution
 \begin{eq}{l}
\int {d^4q\over\pi}\de(q^2-m^2)e^{iqx}=
-{\p\over\p {x^2\over 4}}
 \cl N_0(\sqrt{m^2x^2\over4})
\end{eq}and the causally supported  Fourier transforms
 \begin{eq}{rl}

\int {d^4q\over i\pi^2}
{\Ga(2+n )\over (q_{\ro P}^2-m^2)^{2+n }}e^{iqx}
&=\ep(x_0)\int {d^4q\over2\pi}\ep(q_0)\de^{(1+n)}(m^2-q^2)e^{iqx}\cr

&=i\pi({d\over dm^2})^{1+n }
{\p\over\p {x^2\over 4}} \vth(x^2)
\cl E_0({m^2x^2\over4})\cr
&=\left\{\begin{array}{rl}
i\pi[\de({x^2\over4})-\vth(x^2)m^2\cl E_1({m^2x^2\over4})],& n =-1\cr
-i\pi\vth(x^2)(-{x^2\over4})^ n \cl E_ n ({m^2x^2\over4}),
&n =0,1,\dots\end{array}\right.\cr
\end{eq}The Kepler (Yukawa) factor ${1\over r}$-singularity is embedded into the
lightcone Dirac
distribution ${\p\over\p x^2}\vth(x^2)=\de(x^2)$ for the simple pole $n=-1$.

Feynman propagators
of scalar particle fields
come with simple poles.

\subsection{Time and Position Frames}

By partial Fourier transformation with respect to energy and momentum
one obtains the  embedded time and position representations
\begin{eq}{rl}
\pi g(m^2,x)&=\int {d^4q\over\pi}~e^{iqx}~\tilde g(m^2,q)=
  \int {d^3q\over\pi}~ e^{-i\rvec q\rvec x} g(q_0,x_0)\cr\cr
  &=\int dq_0~ e^{iq_0x_0}[
  \vth(q_0^2-m^2) g^c(|\rvec q|,\rvec x)
 +\vth(m^2-q_0^2) g ^{nc}(i|\rvec q|,\rvec x)]\cr
\end{eq}

{\scriptsize
\begin{eq}{c}
\begin{array}{|c||c||c|c|}\hline
&\hbox{\bf time}
&\hbox{\bf position}
&\hbox{\bf position}\cr
&&\hbox{ (compact)}
&\hbox{(noncompact)}\cr\hline
\tilde g(m^2,q)& g(q_0,x_0)& g^c(|\rvec q|,\rvec x)&
 g ^{nc}(i|\rvec q|,\rvec x)\cr\hline
&q_0=\sqrt{m^2+\rvec q^2}&|\rvec q|=\sqrt{q_0^2-m^2}&
i|\rvec q|=|Q|=\sqrt{m^2-q_0^2}\cr
\hline\hline

\hbox{\bf Lorentz scalars}&&&\cr\hline

\hfill\de(m^2-q^2)& {\cos q_0x_0\over q_0}
&2{\sin |\rvec q|r\over r} &0\cr\hline

\hfill
\ep(q_0)\de(m^2-q^2)
& i{\sin q_0x_0\over q_0}
&  2\ep(q_0){\sin |\rvec q|r\over r}
&0\cr\hline

\hfill
\ep(q_0)\de'(m^2-q^2)
&
{x_0^2\over2 q_0}~ij_1(q_0x_0)
& \ep(q_0){\cos |\rvec q|r\over|\rvec q|}
&0\cr\hline

{1 \over i\pi}{1\over q^2_\ro P -m^2}
& \ep(x_0)i{\sin q_0x_0\over q_0}
&\m~2 i{\cos |\rvec q|r\over r}
&\m~ 2i{ e^{-  |Q|r}\over r} \cr\hline

{1 \over i\pi}{1\over (q^2_\ro P -m^2)^2}
&  \ep(x_0)~ {x_0^2\over 2q_0}~ ij_1(q_0x_0)
&-i{\sin |\rvec q|r\over |\rvec q|}
& -i{  e^{-  |Q|r}\over |Q|}\cr\hline\hline

\hbox{\bf Lorentz vectors}&&&\cr\hline

\hfill q\de(m^2-q^2)&
{\scriptsize \pmatrix{
\hfill i\sin q_0x_0\cr
{\rvec q\over q_0}~\cos q_0x_0\cr}}
&2\rvec q^2{\scriptsize\pmatrix{
\hfill{q_0\over|\rvec q|}~ j_0( |\rvec q|r)\cr
{\rvec x\over r}~ij_1(|\rvec q|r)
\cr}}
&0\cr\hline

q\ep(q_0)\de(m^2-q^2)&
{\scriptsize  \pmatrix{\hfill \cos q_0x_0\cr
{\rvec q\over q_0}~i\sin q_0x_0\cr}}
&\ep(q_0)2\rvec q^2{\scriptsize\pmatrix{
\hfill {q_0\over|\rvec q|}~j_0(|\rvec q|r)\cr
{\rvec x\over r}~ij_1(|\rvec q|r)
\cr}}
&0\cr\hline

q\ep(q_0)\de'(m^2-q^2)&
-{x_0^2\over2}{\scriptsize  \pmatrix{\hfill j_0(q_0x_0)\cr
{\rvec q\over q_0}~ij_1(q_0x_0)\cr}}
&\ep(q_0){\scriptsize\pmatrix{
\hfill {q_0\over|\rvec q|}~\cos|\rvec q|r\cr
{\rvec x\over r}~i\sin |\rvec q|r
\cr}}
&0\cr\hline

{1 \over i\pi}{q\over q^2_\ro P -m^2}
& \ep(x_0)
{\scriptsize  \pmatrix{
\hfill  \cos q_0x_0\cr
{\rvec q\over q_0}~i\sin q_0x_0\cr}}
&\m~2\rvec q^2{\scriptsize\pmatrix{
\hfill {q_0\over|\rvec q|}~in_0(|\rvec q|r)\cr
-{\rvec x\over r}~n_1(|\rvec q|r)
\cr}}
&\m~2|Q|^2{\scriptsize\pmatrix{
\hfill -i{q_0\over |Q|}~ k_0(|Q|r) \cr
\hfill {\rvec x\over r}~k_1(|Q|r)\cr}}
 \cr\hline

{1 \over i\pi}{q\over (q^2_\ro P -m^2)^2}
& -\ep(x_0) {x_0^2\over2}
{\scriptsize  \pmatrix{
\hfill j_0(q_0x_0)\cr
{\rvec q\over q_0}~ ij_1(q_0x_0)
\cr}}
&\o\hfill-{\scriptsize\pmatrix{
 {q_0\over |\rvec q|}~i\sin  |\rvec q|r\cr
\hfill{\rvec x\over r}~\cos  |\rvec q|r\cr}}
&\o\hfill -{\scriptsize\pmatrix{
\hfill i{q_0\over |Q|} \cr {\rvec x\over r}\cr}}
e^{-  |Q|r}
 \cr\hline

\end{array}\cr\cr
\hbox{\bf Time and Position Representations for $\R^4$-Spacetime}
\end{eq}
}

\noindent with - for higher order poles
\begin{eq}{l}
{\p\over\p |m|}=2|m|{\p\over\p m^2}\cong
{|m|\over q_0}{\p\over \p q_0}
\cong-{|m|\over |\rvec q|}{\p\over \p |\rvec q|}
\cong  {|m|\over |Q|}{\p\over \p |Q|}\cr
\end{eq}There arise the scalar
 and vector   Bessel functions $j_{L=0,1}$
(spherical waves for free particles), Neumann functions $n_L$ and
Macdonald functions $k_L$ (for Yukawa interactions and forces).
$r=0$-singular and $r=0$-ambiguous
 elements  ${\rvec x\over r}$ which are no position representations
come with simple and double  poles and
are marked with $\m$ and $\o$ resp.

With the embedding $\R^2\inmap \R^4$ the  time representations
$\R\inmap\R$ remain  simple
poles, the position representation $\R\inmap\R^3$ come as scalar dipoles
and vectorial tripoles as seen
in the projections for trivial momenta $\rvec q=0$ (time frame)
and trivial
energy $q_0=0$ (position frame with `imaginary' momenta) resp.

{\scriptsize
\begin{eq}{c}
\begin{array}{|c||c|c|}\hline
&\hbox{\bf time frame}
&\hbox{\bf position frame }\cr
&x_0=t&\cr
\hline
\tilde g(m^2,q)& g(|m|,t)&
 g ^{nc}(i|m|,\rvec x)\cr
&=\int{d^3x\over 8\pi} g(m^2,x)&
=\int{dx_0\over 2} g(m^2,x)\cr
\hline
&(q_0,\rvec q)=(|m|,0)&(q_0,|\rvec q|)=(0,i|m|)\cr
\hline\hline

\hbox{\bf Lorentz scalars}&&\cr\hline

\hfill\de(m^2-q^2)&{\cos mt\over |m|}
&0\cr\hline

\hfill \ep(q_0)\de(m^2-q^2)
&i{\sin mt\over m}
&0\cr\hline

{1 \over i\pi}{1\over q^2_\ro P -m^2}
&\ep(t)i{\sin mt\over m}
&\m~ 2i{ e^{-  |m|r}\over r} \cr\hline

{1 \over i\pi}{1\over (q^2_\ro P -m^2)^2}
& \ep(t)~i{\sin |m|t-|m|t\cos m t\over 2|m|^3}
&- i {e^{-  |m|r}\over |m|} \cr\hline\hline

\hbox{\bf Lorentz vectors}&&\cr\hline

\phantom{\ep(q_0)}q\de(m^2-q^2)&
{\scriptsize \pmatrix{
\hfill i\sin |m|t\cr
0\cr}}
&0\cr\hline

q\ep(q_0)\de(m^2-q^2)&
{\scriptsize  \pmatrix{\hfill \cos mt\cr
0\cr}}
&0\cr\hline

{1 \over i\pi}{q\over q^2_\ro P -m^2}
& \ep(t)
{\scriptsize  \pmatrix{
\hfill  \cos mt\cr
0\cr}}
&\m~{\scriptsize\pmatrix{
0\cr 2{\rvec x\over r}~{1+|m|r\over r^2}e^{-|m|r}\cr}}
 \cr\hline

{1 \over i\pi}{q\over (q^2_\ro P -m^2)^2}
& \ep(t)
{\scriptsize  \pmatrix{
\hfill -{t \sin |m|t\over 2|m|}\cr
0
\cr}}
&\o{\scriptsize\pmatrix{
\hfill 0 \cr -{\rvec x\over r}\cr}}
e^{-  |m|r}
 \cr\hline

{1 \over i\pi}{2q\over (q^2_\ro P -m^2)^3}
& \ep(t)
{\scriptsize  \pmatrix{
\hfill {t( \sin |mt|-|mt|\cos mt)\over 4|m|^3}\cr
0
\cr}}
&{\scriptsize\pmatrix{
\hfill 0 \cr \rvec x\cr}}
{e^{-  |m|r}\over 2|m|}
 \cr\hline

\end{array}
\cr\cr
\hbox{\bf Time and Position Projection for $\R^4\cong\R\pl\R^3$}

\end{eq}
}

\noindent The time representations from the Dirac and principal value
distributions
have nildimensions $N=0,1,2$ for poles,  dipoles, tripoles.
The $r=0$ regular  nonambiguous
position representation matrix elements
are  the knotless Kepler bound state wave functions  above,
embedded  into the principal value ener\-gy-mo\-men\-tum
distributions for spacetime representations with timelike support $x^2>0$
\begin{eq}{rcl}
\rstate{1,\rvec 0}
\sim &e^{-|m|r}&\inmap
\int {d^4q \over\pi^2}{|m|\over (q^2_\ro P -m^2)^2}e^{iqx}\cr
\rstate{2,\rvec 1}
\sim &2|m|\rvec x~ e^{-|m|r}&\inmap
\int {d^4q \over i\pi^2}{4m^2 q\over (q^2_\ro P -m^2)^3}e^{iqx}\cr
\end{eq}

The   complex representation  functions for 4-di\-men\-sio\-nal spacetime, e.g.
\begin{eq}{l}
\ol\C^2\x\Om^2\ni q\mape {q \over q^2-m^2}\in\ol\C^2\x\Om^2
\end{eq}give as  energy and momentum projected residues
for the Lorentz scalar functions
\begin{eq}{lcrrl}
\Res{\pm |m|}{1\over q^2-m^2}
&=&\oint _{\pm |m|}{d^4q\over 2i\pi^2}\pi\de(\rvec q){1\over q^2-m^2}
=&\oint_{\pm |m|} {dq\over 2i\pi}{1\over q^2-m^2}&=\pm {1\over 2|m|}\cr
\Res{\pm i|m|}{1\over q^2-m^2}&=&
\oint_{\pm i|m|} {d^4q\over 2i\pi^2}\de(q_0){1\over q^2-m^2}
=&-\oint_{\pm i|m|} {d^3q\over 2i\pi^2}{1\over q^2+m^2}&=\mp{i|m|\over 2}\cr
\end{eq}and for the Lorentz vector $q=q_0\bl 1_2+ \rvec q$
a trace  residue $\tr q=2q_0$ for the
energy projection
\begin{eq}{l}
\tr \Res{\pm |m|}{q\over q^2-m^2}
=\oint_{\pm |m|} {dq\over 2i\pi}{ 2q^3\over q^2-m^2}=m^2\cr
\end{eq}

\subsection
[Residual Products (Feynman Integrals)]
{Residual  Products (Feynman Integrals)}

Pointwise  products of Feynman  propagators
convolute energy-mo\-men\-tum distributions
which, in general however,
are not convolutable. For particle propagators,
there arise undefined local products
(`divergencies') of generalized functions
 from the  imaginary principal value
 for the causally supported part
 ${1 \over i\pi}{1\over q^2_\ro P -m^2}\sim i\pi\de(x^2)+\dots$
\begin{eq}{rccc}
   &
   [-{1\over x^2_\ro P}+i\pi\de(x^2)+\dots]
   &\m&[-{1\over x^2_\ro P}+i\pi\de(x^2)+\dots]\cr
\sim&[\de(q^2 -m_1^2)+ {1 \over i\pi}{1\over q^2_\ro P -m_1^2}]&*&
[\de(q^2 -m_2^2)+ {1 \over i\pi}{1\over q^2_\ro P -m_2^2}]
\cr

\end{eq}

The convolution of
 two Feynman  distributions
\begin{eq}{l}
\mp{1\over i\pi^2}
{
{\scriptsize\pmatrix{1\cr q\cr}}
\Ga(2+ n_1 )\over (q^2\mp io -m_1^2)^{2+ n_1} }
*\mp{1\over i\pi^2}{{\scriptsize\pmatrix{1\cr q\cr}}
\Ga(2+ n_2 )\over (q^2\mp io -m_1^2)^{2+ n_2} }\cr
\hskip5mm ={1\over i\pi^2}\int_0^1d\ze_{1,2}\de(\ze_1+\ze_2-1)\int {d^4p\over i\pi^2}
{
{\scriptsize\pmatrix{1&q\ze_1\cr q\ze_2
&-p\ox p+ q\ox q\ze_1\ze_2\cr}}
\ze_1^{1+ n_1}\ze_2^{1+ n_2} \Ga(4+ n_1+ n_2)
\over\brack{
 p^2\mp io+q^2\ze_1\ze_2  -m_1^2\ze_1 -m_2^2\ze_2 }^{4+ n_1+ n_2}   }\cr
\end{eq}involves the tensor $p\ox p- q\ox q\ze_1\ze_2\then
{p^2\over 4}\bl1_4- q\ox q\ze_1\ze_2$  for the  vector-vector
convolution.
Taking the $q$-dependent residue of
the  relative ener\-gy-mo\-men\-ta
\begin{eq}{l}
\mp\int{d^4p\over i\pi^2}{\Ga(3+n)\over (p^2\mp io+a)^{3+n}}
=\mp{1\over 2}\int{d^4p\over i\pi^2}{p^2~\Ga(4+n)\over (p^2\mp io+a)^{4+n}}
={\Ga(1+n)\over (\mp io+a)^{1+n}}\cr
\end{eq}and with the  notation for the  different contours
\begin{eq}{l}
\SO_0(1,3)\sx\R^4:~~
({\stackrel{\rm R}*},q^2)=\left\{\begin{array}{ll}
(\mp{*\over i\pi^2},q^2\mp io),&\hbox{Feynman}\cr
(\mp{*\over 2i\pi^2},(q\mp io)^2),&\hbox{causal}\cr\end{array}\right.
\end{eq}the residual  scalar-scalar,
vector-scalar and vector-vector product reads
\begin{eq}{l}
\R^4:\left\{\begin{array}{rcll}
{\Ga(2+ n_1 )\over(q^2-m_1^2)^{2+ n_1}}
&{\stackrel{\rm R}*}&
{\Ga(2+ n_2) \over(q^2-m_2^2)^{2+ n_2}}
&=\int_0^1 d\ze  {\ze^{1+ n_1}(1-\ze)^{1+ n_2} \Ga(2+ n_1+ n_2)
\over[q^2\ze(1-\ze)-m_1^2\ze -m_2^2(1-\ze)]^{2+ n_1+ n_2} }\cr
{q~\Ga(2+ n_1 )
\over (q^2-m_1^2)^{2+n_1}}
&{\stackrel{\rm R}*}&{\Ga(2+ n_2 )\over ( q^2-m_2^2)^{2+n_2}}&=
\int_0^1  d\ze{q~\ze^{1+n_1}(1-\ze)^{2+n_2}\Ga(2+n_1+n_2)
 \over [q^2\ze(1-\ze)-m_1^2\ze -m_2^2(1-\ze)]^{2+n_1+n_2}}\cr
{q~\Ga(2+ n_1 )
\over (q^2-m_1^2)^{2+n_1}}
&{\stackrel{\rm R}*}&
{q~\Ga(2+ n_2 ) \over ( q^2-m_2^2)^{2+n_2}}&=
\int_0^1  d\ze{-({1\over2}\bl1_4+q\ox q{\p\over\p q^2})\ze^{1+n_1}(1-\ze)^{1+n_2}\Ga(1+n_1+n_2)
 \over [q^2\ze(1-\ze)-m_1^2\ze -m_2^2(1-\ze)]^{1+n_1+n_2}}\cr
\end{array}\right.\end{eq}

The $q^2$-poles in the residual products for the energy and momentum
rational complex functions
disappear in the residual product of the
energy-momentum pole functions
\begin{eq}{rlr}
{q \over q^2}~~
{\stackrel{\rm R}*}~~
{2q \over ( q^2-m^2)^3}&=-({1\over2}\bl1_4+q\ox q{\p\over\p q^2})
&{1\over q^2}\int_0^1  d\ze{1-\ze \over \ze -{m^2\over q^2}}\cr
{q \over (q^2)^2}~~
{\stackrel{\rm R}*}~~
{2q \over ( q^2-m^2)^3}&=-({1\over2}\bl1_4+q\ox q{\p\over\p q^2})
&{1\over (q^2)^2}\int_0^1  d\ze{\ze
 \over (\ze -{m^2\over q^2})^2}\cr
\end{eq}with the residue sum in the closed complex plane
(there is a nontrivial residue at the holomorphic point $\ze=\infty$), e.g.
\begin{eq}{l}
M^2({m^2\over q^2})
=-\int_{0}^{1}
d\ze~{1-\ze\over \ze-{m^2\over q^2}}=-\sum\Res{}{1-\ze\over \ze-{m^2\over q^2}}
\log{\ze-1 \over \ze}
=1-(1-{m^2\over q^2})\log(1- {q^2\over m^2})\cr
\end{eq}

\section{Lorentz Compatible Spin Embedding}

The embedding of position  representations into  Minkowski spacetime
has to embed the harmonic momentum polynomials $(\rvec q)^{2J}=|\rvec q|^{2J}
\ro Y^{2J}(\phi,\th)$ and has to interpret this embedding
with respect to  time representations involved.

The connection
between spin $\SO(3)$ and its covering $\SU(2)$
  to the  Lorentz group    $\SO_0(1,3)$ with its covering $\SL(\C^2)$
  is given by transmutators
as  representatives of the symmetric space
  $\SL(\C^2)/\SU(2)\cong\SO_0(1,3)/\SO(3)$, i.e. of the orientation manifold of
  the   spin group.
 All those transmutators (boost representations) are products of
  the two $(2\x 2)$ transmutators from Pauli spinors
$V\cong \C^2$ to left and right handed Weyl
spinor $V_L\cong\C^2\cong V_R$
\begin{eq}{l}
s(\ul q):V\mape V_L,~~\hat s(\ul q):V\map V_R,~~
\hat s=s^{-1\star}
\end{eq}pa\-ra\-met\-rizable with normalized positive ener\-gy-mo\-men\-ta
\begin{eq}{rl}
q^2> 0,~~{q\over \sqrt{q^2}}=\ul q&=\ul q_0\bl1_2+\rvec{\ul q}
={\scriptsize\pmatrix{\ul q_0+\ul q_3&\ul q_1-i\ul q_2\cr
 \ul q_1+i\ul q_2&\ul q_0-\ul q_3\cr}}\cr
\ul{\d q}&=\ul q_0\bl1_2-\rvec{\ul q}
,~~\ul q^2=1=\ul{\d q}^2\cr
\end{eq}Both Weyl transmutators
embed the unit $\bl1_2$ for the Pauli spinor space
and the spherical harmonics $\ro Y^1(\phi,\th)={\rvec q\over |\rvec q|}$
into the normalized ener\-gy-mo\-men\-ta
\begin{eq}{rl}
s(\ul q)\bl1_2s^\star(\ul q)&=\ul q,~~
\hat s(\ul q)\bl1_2\hat s^\star(\ul q)=\ul {\d q}, ~~
s(\ul q),\hat s(\ul q)=s(\ul{\d q})\in\SL(\C^2)\cr
\then s(\ul q)
&=u({\rvec q\over |\rvec q|})
\o e^{{\be\over2}\si_3}\o
u^\star({\rvec q\over |\rvec q|}),~~
\tanh\be={q_0\over|\rvec q|},~~
u({\rvec q\over |\rvec q|})\in\SU(2)\cr
&={1\over\sqrt{2 (1+\ul q_0)}}
{\scriptsize\pmatrix{1+\ul q_0+\ul q_3&\ul q_1-i\ul q_2\cr
 \ul q_1+i\ul q_2&1+\ul q_0-\ul q_3\cr}}\cr

\end{eq}

Now the general case:
An $\SU(2)$-representation $[2J]$ with spin $J=0,{1\over2},\dots$
is embedded into
finite dimensional irreducible representations
$[2L|2R]$ with left and right `spin' $L,R$ of the Lorentz group
$\SL(\C^2)$ for
\begin{eq}{l}
[2J]\inmap [2L|2R]\hbox{ for }\left\{\begin{array}{l}
L+R\ge J\cr
L+R+J\hbox{ integer}\end{array}\right.
\end{eq}with the $\SU(2)$-decomposition\begin{eq}{l}
[2L|2R]\cong{\PL_{J=|L-R|}^{L+R}}[2J],~~\C^{(1+2L)(1+2R)}\cong{\PL_{J=|L-R|}^{L+R}}\C^{1+2J}
\end{eq}The Lorentz group acts upon
the totally symmetrized products
${\OD^{2L}}V_L\ox {\OD^{2R}}V_R\cong\C^{(1+2L)(1+2R)}$
of  Weyl spaces.
The  transmutators
\begin{eq}{l}
s^{[2L|2R]}(\ul q)={\OD^{2L}}s(\ul q)\ox {\OD^{2R}}\hat s (\ul q):
{\OD^{2L}}V\ox {\OD^{2R}}V\map {\OD^{2L}}V_L\ox {\OD^{2R}}V_R\cr
\end{eq}allow the Lorentz compatible embedding of spin properties.

E.g., the Minkowski representation of the boosts
\begin{eq}{l}
s^{[1|1]}(\ul q):
V\ox V\map V_L\ox V_R\cr
s^{[1|1]}(\ul q)=s(\ul q)\ox \hat s(\ul q)=\La( \ul q)={\scriptsize\pmatrix{
\ul q_0&\rvec{\ul q}\cr \rvec{\ul  q}&
\bl 1_3+{\rvec {\ul q}\ox\rvec{\ul q}\over1+\ul q_0}\cr}}
\in \SO_0(1,3)
\end{eq}gives the Lorentz compatible embeddings with
the projectors for spin 0 and 1
\begin{eq}{l}
V_L\ox V_R\cong\C^4=\C\pl\C^3,\left\{\begin{array}{cl}

\La( \ul q){\scriptsize\left(\begin{array}{c|c}
1&0\cr\hline 0&0\cr\end{array}\right)}\La^{-1}( \ul q)&=
\ul q\ox \ul {\d q}\cong   {q^j q_k\over q^2}\cr
\La( \ul q){\scriptsize\left(\begin{array}{c|c}
0&0\cr\hline 0&\bl1_3\cr\end{array}\right)}\La^{-1}( \ul q)
&=\bl 1_4-\ul q\ox \ul{\d q}\cong \de_k^j -{q^jq_k\over q^2}
\end{array}\right.
\end{eq}

This example is characteristic:
The totally symmetric spherical harmonics
are embedded for integer spin in symmetric $\SO_0(1,3)$-representations
\begin{eq}{rl}
J=0,1,\dots:~~
[2J]&\inmap[2L|2L]\hbox{ with }2L\ge J\cr
({\rvec q\over |\rvec q|})^{2J}
&\inmap(\ul q)^{4L}_{2J}
= {\OD^{2L}}\ul q\stackrel{2J}\ox {\OD^{2L}}\ul{\d q}\cr
\end{eq}with the
decomposition of the  unit matrix into projectors
\begin{eq}{rl}
J=0,1,\dots:~~
\bl 1_{(1+2L)^2}={\PL_{J=0}^{2L}} (\ul q)^{4L}_{2J},~~
(\ul q)^{4L}_{2J}
&= s^{[2L|2L]}(\ul q)\bl 1_{1+2J} s^{-1[2L|2L]}(\ul q)\cr
(\ul q)^{4L}_{2J} \o (\ul q)^{4L}_{2J'}
&=\de_{JJ'}(\ul q)^{4L}_{2J}\cr

\end{eq}

In generalization of the two Weyl representations
there arise two embedding types
for half-integer spin, conjugated
to each other. They can be Clebsch-Gordan composed from the
two Weyl transmutators
\begin{eq}{rl}
J={1\over2},{1\over 3}\dots:~~
[2J]&\inmap \left\{\begin{array}{l}
[1+2L|2L]\cr
[2L|1+2L]\cr\end{array}\right.
\hbox{ with }2L\ge J-{1\over 2}\cr
({\rvec q\over |\rvec q|})^{2J}
&\inmap\left\{\begin{array}{rcrcl}
(\ul q)^{1+4L}_{2J}&=&
 {\OD^{1+2L}}\ul q&\stackrel{2J}\ox& {\OD^{2L}}\ul{\d q}\cr
(\ul{\d q})^{1+4L}_{2J}&=&
 {\OD^{2L}}\ul q&\stackrel{2J}\ox& {\OD^{1+2L}}\ul{\d q}\end{array}\right.
\end{eq}

An appropriate $\D(1)$-dilatation factor
gives transmutators from $\U(2)$ to  $\GL(\C^2)$,
 i.e. representatives of the
symmetric space $\GL(\C^2)/\U(2)$
\begin{eq}{rl}
q^2\ge0:~~s(q)=\sqrt{q^2}s(\ul q)&=
u({\rvec q\over |\rvec q|})
\o \sqrt{q^2}~e^{{\be\over2}\si_3}\o
u^\star({\rvec q\over |\rvec q|})\in\GL(\C^2)\cr
\end{eq}Therewith the
harmonic polynomials are Lorentz compatibly embedded
\begin{eq}{l}
({\rvec q\over |\rvec q|})^{2J}\inmap (\ul q)_{2J}^{K},~~
(\rvec q)^{2J}\inmap (q)_{2J}^{N}=(\sqrt{q^2})^{2K}(\ul q)^{N}_{2J}
\end{eq}with the examples from above
  for Lorentz scalar, left and right Weyl spinor and
Lorentz vector (with the projectors $\bl1_4=\bl 1_1+\bl1_3$)
\begin{eq}{ll}
(\rvec q)^0\inmap ( q)_0^{0} =1&
(\rvec q)^1\inmap\left\{\begin{array}{rll}
(q)_{1}^{1} &= q
&=s(q)\bl1_2 s^\star(q)\cr
( \d q)_{1}^{1} &= {\d q}&=\hat s(q)\bl1_2 \hat s^\star(q)\end{array}\right.\cr
(\rvec q)^0\inmap(q)^{2}_{0} = q\ox \d q=\La(q)\bl 1_1\La(q)^{-1},&
(\rvec q)^2\inmap( q)_2^{2}=q^2\bl 1_4- q\ox \d q
=\La(q)\bl 1_3\La(q)^{-1}\cr
\end{eq}

Convolutions of ener\-gy-mo\-men\-ta are understood to involve also the
tensor products of the spin representations. E.g.,
in the vector-vector convolution  above
there arises the projectors for spin 0 and 1
\begin{eq}{l}
q*\d q\then {1\over 2}\bl 1_4 +q\ox \d q{\p\over\p q^2}
={1\over 2}
(\ul q)^2_2 +[{1\over 2} + q^2 {\p\over\p q^2}]
(\ul q)^2_0
\end{eq}

\section
[Residual Representations of Future Cones]
{Residual Representations\\of Future Cones}

Causal (advanced and retarded) and Feynman
multipole ener\-gy-mo\-men\-tum  distributions
lead  - via their Fourier transforms with appropriate integration contours
  - to representation
matrix elements of different
symmetric spaces - of the causal bicone (future and past  cone)
and of the tangent spacetime translations resp.
Feynman distributions with $\de(q^2-m^2)$ from a simple pole represent
 spacetime translations
 as inhomogeneous subgroup of
 irreducible unitary Poincar\'e group representations,
  acting on free particles.
The representations of the future cone
as model of nonlinear spacetime\cite{S991,S011}
involves  higher order energy-momentum poles. They are no particle propagators. They
will be used to  determine the masses and normalization
of particles for the construction of  Feynman propagators.

\subsection{Spacetime Future Cones}

1-di\-men\-sio\-nal time future is embedded into the future cones of
2- and 4-di\-men\-sio\-nal Minkowski spacetime
\begin{eq}{cccl}
\R_\od\ni t_\od =\vth(t)t&\inmap &
\vth(x^0)\vth(x^2)
{\scriptsize\pmatrix{x^ 0+x^3&0\cr0&x^0- x^3\cr}}&=x_\od\in \R^2_\od\cr
&\inmap&
\vth(x^0)\vth(x^2){\scriptsize\pmatrix{x^0+x^3&x^1-ix^2\cr
x^1+ix^2&x^0- x^3\cr}}&=x_\od\in \R^4_\od
\end{eq}with associated orthochronous groups - trivial, abelian, simple
\begin{eq}{l}
\{1\}=\SO(1)\inmap\SO_0(1,1)\inmap\SO_0(1,3)\cr
\end{eq}

Time future
is the causal group $\D(1)=\exp \R$
\begin{eq}{rl}
\R_\od\ni t_\od&=e^{\psi^0}\in\D(1)\cr
\R_\od&\cong\D(1)\cong\GL(\C)/\U(1)\cr
\end{eq}The 2-di\-men\-sio\-nal  future cone
 is the direct product of causal group and selfdual
 Lorentz dilatation group
 \begin{eq}{rl}
\R_\od^2\ni x_\od=
{\scriptsize\pmatrix{x_\od^ 0+x^3&0\cr0&x_\od^0- x^3\cr}}
&=e^{\psi^0+\si_3\psi^3}
\hbox{ with}\left\{\begin{array}{rl}
x_\od^2&=e^{2\psi^0}\cr
 {x_\od^ 0+x^3\over x_\od^ 0-x^3}&=e^{2\psi^3}\end{array}\right.\cr
\R^2_\od&\cong\D(1)\x\SO_0(1,1)\cr
\end{eq}The 4-di\-men\-sio\-nal  future cone
is a   homogeneous space with
2-di\-men\-sio\-nal future $\R^2_\od$ as abelian Cartan substructure
\begin{eq}{rl}
\R_\od^4\ni x_\od&={\scriptsize\pmatrix{x_\od^0+x^3&x^1-ix^2\cr
x^1+ix^2&x_\od^ 0- x^3\cr}}
=e^{\psi^0+\rvec\psi}=u({\rvec x\over r})\o
e^{\psi^0+\si_3 |\psi|}\o u({\rvec x\over r})^\star\cr
&\hfill\hbox{with }\left\{\begin{array}{rlrl}
 x_\od^2&= e^{2\psi^0},& {x_\od^ 0+r\over x_\od^ 0-r}&=e^{2|\rvec\psi|}\cr
 {\rvec \psi\over|\rvec \psi|}&={\rvec x\over r},&
u({\rvec x\over r })&\in\SU(2)\end{array}\right.\cr

\R^4_\od&\cong\D(1)\x\SO_0(1,3)/\SO(3)\cr
&\cong\D(1)\x\SO_0(1,1)\x\Om^2
\cong\GL(\C^2)/\U(2)\cr

\end{eq}

The cones as irreducible orbits of $\D(1)\x\SO_0(1,s)$, $s=0,1,3$,
 are
used as  strict futures, open
without `skin', i.e. without the strict presence $x=0$ and without  lightlike
translations for nontrivial position $s=1,3$
\begin{eq}{l}
x_\od\in\R_\od^{1+s}\then x_\od^2>0
\end{eq}

1-di\-men\-sio\-nal
and 4-di\-men\-sio\-nal future
are the first two entries
of the symmetric space chain $\GL(\C^n)/\U(n)$, $n=1,2,\dots$,
which are the
manifolds of the unitary groups in the general linear group,
canonically pa\-ra\-met\-rized in the polar decomposition $g=u\o |g|$
with the real $n^2$-di\-men\-sio\-nal ordered absolute values $x_\od=|g|=\sqrt{g^\star\o g}
\in\R^{n^2}_\od$
of the general linear group. They are the positive cone of the
ordered $C^*$-algebras with the complex $n\x n$-matrices.

In residual representations the future
cone $\R_\od^{1+s}=G/H$ is
canonically pa\-ra\-met\-rized
by  translations  which constitute
the tangent space $\log G/H$ of the future cone
\begin{eq}{l}
\R^{1+s}
\cong\left\{\begin{array}{ll}
\log\D(1),&s=0\cr
\log \D(1)\pl\log\SO_0(1,1),&s=1\cr
\log \GL(\C^2)/\U(2),&s=3\cr
\end{array}\right.
\end{eq}The cone is embedded into its tangent
space. The future cone $\R^4_\od\cong\GL(\C^2)/\U(2)$
as the orientation manifold
 of unitary groups  is taken as model for
nonlinear spacetime. The
$\GL(\C^2)$-action by left multiplication involves the external
Lorentz group. The   group
$\U(2)$ of the equivalence classes is
used for internal degrees of freedom (hyperisospin).
The related structures\cite{S982,S011}  will not
be considered in more detail in the following.

\subsection{Residual Representations of Time Future}

The residual representations of time future
by the advanced energy distributions are characterized by one
 compact invariant and nildimension $N$
\begin{eq}{l}
{1\over2i\pi}{\Ga(1+ N  ) \over ( q -io-m )^{1+ N  }}
={1\over 2}[ \de^{( N  )}(m- q )
+{1\over i\pi}{\Ga(1+ N  )  \over ( q_\ro P-m)^{1+ N  }}]\cr
\end{eq}They are
representation  matrix elements of the causal group
$\D(1)$
\begin{eq}{l}
\R_\od\ni t_\od \mape

\int {dq\over 2i\pi} {\Ga(1+ N  )  \over ( q - io -m)^{1+ N  }}e^{iqt }
=(it_\od)^ N   e^{imt_\od}\cr
\end{eq}

\subsection
[Residual Representations of 2-Dimensional Future]
{Residual Representations\\of 2-Dimensional Future}

The residual representations of 2-di\-men\-sio\-nal future
will be constructed from the advanced ener\-gy-mo\-men\-tum
distributions
 \begin{eq}{l}
  {1 \over 2i\pi}
 {1\over (q-io)^2 -m^2}=
 {1\over2}\Bigl[\ep(q_0)\de(q^2 -m^2)+ {1 \over i\pi}
{1\over q^2_\ro P -m^2}\Bigr]\cr
\end{eq}With the  Fourier transforms and their partial projections
one obtains for the representations of time
future and position
\begin{eq}{rl}
\pi g(m^2,x)&=\int d^2q~e^{iqx}~\tilde g(m^2,q)=
  \int dq_3~ e^{-i q_3x^3} g(q_0,x^0)\cr\cr
&  =\int dq_0~ e^{iq_0x^0}[
  \vth(q_0^2-m^2) g^c(q_3,x^3)+\vth(m^2-q_0^2) g^{nc}
  (iq_3,x^3)]\cr
\end{eq}

{\scriptsize

\begin{eq}{c}
\begin{array}{|c||c||c|c|}\hline
&\hbox{\bf spacetime future}
&\hbox{\bf time frame}
&\hbox{\bf position frame}\cr
& x_\od=\vth(x^0)\vth(x^2) x&
x^0_\od= t_\od
&x^3=z\cr
\hline
\tilde g(m^2,q)
& g(m^2,x)
& g(|m|,t_\od)\in\D(1)
&
 g^{nc}(i|m|,z)\in\SO_0(1,1)\cr\hline
&
&(q_0, q_3)=(|m|,0)&(q_0, q_3)=(0,i|m|)\cr
\hline\hline

{1 \over 2\pi}{1\over (q-io)^2 -m^2}&
\hfill -\cl E_0({m^2x_\od^2\over4})
& -{\sin |m|t_\od\over |m|}
&  {e^{-  |mz|}\over|m|} \cr\hline

{1 \over 2i\pi}{q\over (q-io)^2 -m^2}&
\m\hfill
\p_\od\cl E_0({m^2x_\od^2\over4})
&
\cos mt_\od
&\ep(z)e^{-  |mz|}
 \cr\hline
{1 \over 2i\pi}{ q\over ((q-io)^2 -m^2)^2}&
\hfill
-{x_\od\over2}\cl E_0({m^2x_\od^2\over4})
&
-{t_\od\sin |m|t_\od\over 2|m|}
&
 -z{e^{-  |mz|}\over 2|m|}
 \cr\hline

\end{array}\cr\cr
\hbox{\bf Representations of Spacetime Future
$\R^2_\od\cong\R_\od\pl\R$}\cr
\cr\begin{array}{rl}
{\p\over\p{x^2\over 4}}\cl E_0({m^2x_\od^2\over4})
&=\de({x_\od^2\over4})
-m^2\cl E_1({m^2x_\od^2\over4})~~
\hbox{ with }\de(x_\od^2)=\vth(x_0)\de(x^2)\cr\cr
{\p\over\p m^2}{\p\over\p{x^2\over 4}}\cl E_0({m^2x_\od^2\over4})
&=-\cl E_0({m^2x_\od^2\over4}),~~\p_\od={\p\over\p x_\od}=2x_\od{\p\over\p x^2}\cr
\end{array}\cr

\end{eq}
}

With $t_\od=\vth(t)t={1+\ep(t)\over2}t$ the time future projections,
i.e. the representation matrix elements  of the causal group $\D(1)$,
are combined from Dirac and principal value contribution.
The position space projections,
i.e. representation matrix elements  of the orthochronous group $\SO_0(1,1)$,
come from the principal value  only.

Spacetime future representation matrix elements
have to be functions, i.e.
the Dirac distribution $\de(x_\od^2)$ on the forward lightcone
in the Lorentz vector gives
 no  representations, marked by $\m$.
The future lightlike translations  ${\bl1_2\pm\si_3\over 2}x_\od^0$
are  no elements of strict  future $x_\od ^2>0$.

2-di\-men\-sio\-nal future is the {\it rank 2 real Lie group}
$\D(1)\x\SO_0(1,1)$. The residual representations of
these two noncompact groups will be characterized by
two invariants for the characters, both from a continuous spectrum.
Therefore the dipole
in the residual representation will be  supported by
{\it two  Lorentz invariants} for
the hyperbolic-spherical  singularity surface
with the pole function
\begin{eq}{l}
{1\over q^2-m_0^2}-{1\over q^2-m_3^2}
={m_0^2-m_3^2\over (q^2-m_0^2)(q^2-m_3^2)}
=\int_{m_3^2}^{m_0^2}dm^2~{1\over (q^2-m^2)^2}
\end{eq}By the Lorentz compatible embedding
with  tangent $\R^2$-translations and energy-momenta
both invariants contribute to  representations
of the time group $\D(1)$ and the position
space $\SO_0(1,1)$.

On the lightcone $x^2=0$, where
time and position translations coincide $x^3=\pm x^0$,
the contributions  from both invariants  cancel each other
as seen for the vector representation
\begin{eq}{l}
\begin{array}{rl}
\hbox{spacetime future: }\R_\od^2\ni x_\od\mape&
\int_{m_3^2}^{m_0^2}dm^2~
\int {d^2q \over 2i\pi}~{q\over ((q-io)^2 -m^2)^2}
e^{iqx}\cr
&=-{x_\od\over2}\pi[m_0^2\cl E_1({m_0^2x_\od^2\over4})
-m_3^2\cl E_1({m_3^2x_\od^2\over4})]\end{array}\cr
\end{eq}with the projection $x_\od=t_\od\bl1_2+z\si_3$ on time future and position
\begin{eq}{lrl}
\hbox{time future:}&\R_\od\ni t_\od\mape&
\int_{m_3^2}^{m_0^2}dm^2\int {dq \over 2i\pi}~{q\over ((q-io)^2 -m^2)^2}
e^{iqt}\cr
&&=\cos m_0t_\od-\cos m_3t_\od\cr
\hbox{position:}&\R\ni z\mape& \int_{m_3^2}^{m_0^2}dm^2~
\int {dq \over 2i\pi}~{q\over (q^2+m^2)^2}
e^{-iqz}\cr
&&=\ep(z){e^{-  |m_0z|}-e^{-  |m_3z|}\over2}
 \end{eq}

The energy  projected trace residues of the representation functions
are
\begin{eq}{l}
\tr \Res{\mu}
{m_0^2-m_3^2\over (q^2-m_0^2)(q^2-m_3^2)}
=\left\{\begin{array}{rl}
1,&\mu^2=m_0^2\cr
-1,&\mu^2=m_3^2\cr\end{array}\right.\cr
\end{eq}

\subsection
[Residual Representations of 4-Dimensional Future]
{Residual Representations\\of 4-Dimensional Future}

2-di\-men\-sio\-nal future
is a Cartan subgroup of 4-di\-men\-sio\-nal future with
additional 2-sphere degrees of freedom
$\R_\od^4/\R_\od^2\cong\Om^2$.

The  residual representations
of 4-di\-men\-sio\-nal future
by  advanced ener\-gy-mo\-men\-tum  distributions
have as projections to time future  and position
\begin{eq}{rl}
\pi g(m^2,x)&=\int {d^4q\over\pi}~e^{iqx}~\tilde g(m^2,q)=
  \int {d^3q\over\pi}~ e^{-i \rvec q\rvec x} g(q_0,x^0)\cr\cr
&  =\int dq_0~ e^{iq_0x^0}[
  \vth(q_0^2-m^2) g^c(|\rvec q|,\rvec x)
 +\vth(m^2-q_0^2) g^{nc}(i|\rvec q|,\rvec x)]\cr
 \end{eq}

{\scriptsize
\begin{eq}{c}
\begin{array}{|c||c|c|c|}\hline
&\hbox{\bf spacetime future}
&\hbox{\bf time frame}
&\hbox{\bf position frame}\cr
& x_\od=\vth(x^0)\vth(x^2) x&
 x^0_\od=t_\od&\cr
\hline
\tilde g(m^2,q)& g(m^2,x)& g(|m|,t_\od)\in\D(1)&
 g ^{nc}(i|m|,\rvec x)\in\SO_0(1,3)/\SO(3)\cr\hline

&&(q_0,\rvec q)=(|m|,0)&(q_0,|\rvec q|)=(0,i|m|)\cr
\hline\hline

{1 \over \pi}{1\over (q-io)^2 -m^2}
&\m\hfill -
{\p\over\p{x^2\over 4}}
\cl E_0({m^2x_\od^2\over4})
&-{\sin |m|t_\od\over|m|}
& -{ e^{-  |m|r}\over r} \cr\hline

{1 \over \pi}{1\over ((q-io)^2  -m^2)^2}
&\hfill
\cl E_0({m^2x_\od^2\over4})

 &-{\sin|m|t_\od-|m|t_\od\cos mt_\od\over 2|m|^3}
&{e^{-  |m|r}\over2|m|} \cr\hline\hline

{1 \over i\pi}{q\over (q-io)^2 -m^2}
&\m\hfill \p_\od {\p\over\p{x^2\over 4}}
\cl E_0({m^2x_\od^2\over4})
 &\cos mt_\od
&
{\rvec x\over r}~{1+|m|r\over r^2}e^{-|m|r}
 \cr\hline

{1 \over i\pi}{q\over ((q-io)^2 -m^2)^2}
&\m\hfill \p_\od
\cl E_0({m^2x_\od^2\over4})

&
 -{t_\od \sin |m|t_\od\over 2|m|}
 & -{\rvec x\over r}{e^{-  |m|r}\over 2}
 \cr\hline
{1 \over i\pi}{2q\over ((q-io)^2 -m^2)^3}
&\hfill -{x_\od\over2}\cl E_0({m^2x_\od^2\over4})

&
{t_\od( \sin |m|t_\od-|m|t_\od\cos mt_\od)\over 4|m|^3}
& -\rvec x ~{e^{-  |m|r}\over2|m|}
 \cr\hline

\end{array}
\cr\cr
\hbox{\bf Representation of Spacetime Future $\R_\od^4\cong\R_\od\pl\R^3$}
\cr\cr
\begin{array}{rl}
\({\p\over\p{x^2\over 4}}\)^2
\cl E_0({m^2x_\od^2\over4})

&
=\de'({x_\od^2\over4})-m^2\de({x_\od^2\over4})
+m^4\cl E_2({m^2x_\od^2\over4})\cr
\end{array}\cr

\end{eq}
}

4-di\-men\-sio\-nal future is the  real homogeneous space
$\GL(\C^2)/\U(2)$ with rank 2 for a Cartan subgroup $\D(1)\x\SO_0(1,1)$.
Therefore its residual representation will be  supported
by two invariants as
for the 2-di\-men\-sio\-nal case
with a characteristic additional dipole structure\cite{HEI}
to take into account the 2-sphere degrees
of freedom in 3-di\-men\-sio\-nal position
\begin{eq}{l}
{1\over q^2-m_0^2}-{1\over q^2-m_3^2}
-{m_0^2-m_3^2\over (q^2-m_3^2)^2}={(m_0^2-m_3^2)^2\over (q^2-m_0^2)(q^2-m_3^2)^2}

=\int_{m_3^2}^{m_0^2}dm^2(m_0^2-m^2)
~{2\over (q^2-m^2)^3}
\end{eq}Again,
both invariants contribute to  representations
of the time group $\D(1)$ and the symmetric position
space
$\SL(\C^2)/\SU(2)\cong\SO_0(1,1)\x\Om^2\cong\R^3$.

There is one aditional noncompact continuous invariant
 compared with the
one  compact mass
invariant $m^2$
 of the  Poincar\'e group
for free particles as used in the Wigner classification
\begin{eq}{l}
\begin{array}{|c||c|c|c|}\hline
\hbox{rank}&\hbox{Lorentz}&\hbox{Poincar\'e}&
\hbox{expansion}\cr\hline\hline
1&\SO_0(1,1)&\SO_0(1,1)\sx\R^2&\SO_0(2,1)\cr\hline
2&\SO_0(1,3)&\SO_0(1,3)\sx\R^4&\SO_0(2,3)\cr\hline
\end{array}\cr
\end{eq}

With two invariants the vector representations of 4-di\-men\-sio\-nal  future
are
\begin{eq}{l}
\hbox{spacetime future: }
\R_\od^4\ni x_\od\mape
\int_{m_3^2}^{m_0^2}dm^2(m_0^2-m^2)
\int {d^4q \over 2i\pi^2}~{2q\over ((q-io)^2 -m^2)^3}
e^{iqx}\cr
\hfill={x_\od\over2}\pi[
m_0^4\cl E_2({m_0^2x_\od^2\over4})
-m_3^4\cl E_2({m_3^2x_\od^2\over4})
+(m_0^2-m_2^2)m_3^2\cl E_1({m_3^2x_\od^2\over4})
]

\cr
\end{eq}with the projections $x_\od=t_\od\bl1_2+\rvec x$ on time future
and 3-di\-men\-sio\-nal position
\begin{eq}{rl}
\hbox{time future:}&\R_\od\ni t_\od\mape\int_{m_3^2}^{m_0^2}dm^2
(m_0^2-m^2)\int {dq \over 2i\pi}~{q\over ((q-io)^2 -m^2)^3}
e^{iqt}\cr
&\hfill=
\cos m_0t_\od-\cos m_3t_\od+(m_0^2-m_3^2)
{t_\od\sin m_3t_\od\over 2m_3}

\cr
\hbox{position:}&\R^3\ni \rvec x\mape \int_{m_3^2}^{m_0^2}dm^2(m_0^2-m^2)
\int {d^3q \over 2\pi^2}~{i\rvec q\over (\rvec q^2 +m^2)^3}
e^{-i\rvec q\rvec x}\cr
&\hskip10mm=
{\rvec x\over r}[
{1+|m_0|r\over r^2}e^{-  |m_0|r}
-{1+|m_3|r\over r^2}e^{-  |m_3|r}
+(m_0^2-m_3^2){e^{-  |m_3|r}\over 2}]
 \end{eq}

The  energy  projected trace residues of the representation functions
are  as for the Cartan substructure
\begin{eq}{l}
\tr \Res{\mu}
{(m_0^2-m_3^2)^2\over (q^2-m_0^2)(q^2-m_3^2)^2}

=\Res{\mu}[
{2q^3\over q^2-m^2_0}-{2q^3\over q^2-m^2_3}]
=\left\{\begin{array}{rl}
1,&\mu^2=m_0^2\cr
-1,&\mu^2=m_3^2\cr\end{array}\right.
\end{eq}

 A simple  pole ${q\over q^2-m^2}$
has a positive energy projected  residue,
its mass can be associated to a particle.
The related irreducible  time translation representation
with  positive  normalization in an associate
inner product space can be taken over
to define a Feynman propagator
as Fourier transformation of $q\de(q^2-m^2)$ with unitary
representations  $e^{iqx}$ of spacetime translations
by a free particle.
Dipoles ${q\over (q^2-m^2)^2}$
cannot be related to probability valued eigenvectors
for translations, they
come from nondecomposable
2-di\-men\-sio\-nal nondiagonalizable translation representations  with
triagonal nilpotent  Jordan contributions and with a ghost metric\cite{S912,S913}.
Product representations with a dipole can involve poles for particles.
A dipole ${1\over (q^2-m^2)^2}$ has a nontrivial
momentum projected residue.

\section
[Matter as Spacetime Spectrum]
{Matter as Spacetime Spectrum}

\subsection
[Residual Representations of Tangent Spaces]
{Residual Representations of Tangent Spaces}

Complex pole functions of the translation characters
(energy-momenta) $q\mape {Q(q)\over P(q)}$ can
be used both for the representations
of a symmetric space (spacetime)
and for the representations of its tangent space
(spacetime translations).

On a symmetric space  function
$(G/H)_{\rm repr}\ni x\mape  g(x)$
with canonical pa\-ra\-met\-rization,
e.g. $x^Ne^{i m  x}$
 for $\D(1)$ or
$e^{i|m|\rvec x}$ for $\SU(2)$, the tangent space (Lie algebra) action
involves the corresponding derivatives, e.g.
\begin{eq}{l}
{d\over dx}\hbox{ for } \log\D(1)\cong \R,~~
  {\p\over  \p\rvec x}, {\p^2\over  \p\rvec x^2}\hbox{ for }
  \log\SO_0(1,3)/\log \SO(3)\cong \R^3
\end{eq}Therewith a {\it tangent distribution}
of a symmetric space,
e.g. a  Lie algebra distribution for a Lie group,
will be defined by  an inverse derivative with an invariant pole
and a residue $a_{-1}$, familiar
as {\it Green distributions of differential equations}
(in general no functions).
Its Fourier transform defines a complex  tangent representation
function. Tangent distributions come with different integration contours.
In contrast to the normalization of Cartan group representations by
the group unit, the residue of  a tangent representation
has to be determined by another structure (below).

The  causal group time is isomorphic to its tangent space.
Therefore  the tangent  representation functions with appropriate residue
are also group  representation functions
\begin{eq}{l}
\hbox{time }\D(1)\cong \R:~~{a_{-1}\over q-m}
\end{eq}For 3-dimensional position with the rank 2 Euclidean semidirect group  there are two types of  tangent functions
- for integer and half integer spin
\begin{eq}{l}
\hbox{position }\SO(3)\sx \R^3,~~\mu^2=\pm m^2:\left\{\begin{array}{rll}
J=0,1,\dots:&
{a_{-1}(\rvec q)^{2J}\over (\rvec q^2-\mu^2)^{1+J}},&
\hbox{e.g. }{a_{-1}\over \rvec q^2-\mu^2}\cr
J={1\over2},{3\over2},\dots:
&{a_{-1} (\rvec q)^{2J}\over (\rvec q^2-\mu^2)^{{1\over 2}+J}},&
\hbox{e.g. }{a_{-1}\rvec q\over \rvec q^2-\mu^2}\cr
\end{array}
\right.
\end{eq}The Fourier transforms involve the Yukawa potential and force.

The tangent functions for time and position
have to be embedded into Minkowski spacetime tangent
function:
For 2-di\-men\-sio\-nal spacetime one has
with the rank 1 Poincar\'e group
\begin{eq}{l}
\hbox{spacetime }\SO_0(1,1)\sx \R^2:~~
{a_{-1}\over q^2-m^2},~~
{a_{-1}q\over q^2-m^2}
\end{eq}For 4-di\-men\-sio\-nal  spacetime
with the rank 2 Poincar\'e group there are two tangent function types
\begin{eq}{l}
\hbox{spacetime }\SO_0(1,3)\sx\R^4:\left\{\begin{array}{rl}
J=0,1,\dots:&{a_{-1}(q)_{2J}^{4L}\over (q^2-m^2)^{1+2L}}\hbox{ with }2L\ge J\cr
&\hbox{e.g. }{a_{-1}\over  q^2-m^2}\cr
J={1\over2},{3\over2},\dots:&
{ a_{-1}(q)_{2J}^{1+4L}\over (q^2-m^2)^{1+2L}}\hbox{ with }2L\ge J-{1\over 2}\cr
&\hbox{e.g. }{a_{-1}q\over q^2-m^2}\cr
\end{array}\right.
\end{eq}For a given $J$ there are different embeddings $L$, as discussed above.

There  is a decisive difference of tangent distributions  of
position $\R^3$ and
2- and 4-di\-men\-sio\-nal  spacetime $\R^{2,4}$
compared with those of  time $\R$.
In general, tangent distributions $ l\in{\cl G}'$
 are no   symmetric space functions $g\in{\cl G}$, i.e.
 $\cl G'\sup \cl G$.
They are derivatives thereof with respect to the canonical parameters,
e.g. ${e^{-\mu r}\over 2r}=-{d\over dr^2}{e^{-\mu r}\over \mu}$
where the  Yukawa potential arise from the
tangent representation functions
$\{ {1\over (\rvec q^2+\mu^2)^{1+N}}\mid N=0,1,\dots\}$
and the exponential from the symmetric space  representation
functions $\{ {1\over (\rvec q^2+\mu^2)^{2+N}}\mid N=0,1,\dots\}$.
In general, the  tangent representation functions
constitute a vector space only. In contrast to the pointwise multiplicative
 property of symmetric space functions $\cl G\m\cl G\map\cl G$ and
 convolution for their Fourier transforms
   $\tilde{\cl G}*\tilde{\cl G}
 \map\tilde{\cl G}$
the requirement of multiplicative stability for the tangent distributions
does not make sense (`divergencies').
Translations and their representations can be added, but, in general,
they cannot be multiplied. E.g.  a squared Yukawa potential
${e^{-2|m|r}\over r^2}$ does not make sense as a representation. Or,
 Lie algebra representation matrix elements
have no associative multiplicative structure.
However, a tangent vector space (Lie algebra) should be
a module  $\tilde{\cl G}'\in\mod{\tilde{\cl G}}$ with
respect to the residual action with
functions
$\tilde{\cl G}$ for symmetric space (group) representations, i.e.
$\tilde{\cl G}*\tilde{\cl G}'
 \map\tilde{\cl G}'$ and for the Fourier transforms
$\cl G\m\cl G'\map\cl G'$
\begin{eq}{rcll}
\hbox{symmetric space}&*&\hbox{symmetric space}&\map \hbox{symmetric space}\cr
\hbox{symmetric space}&*&\hbox{tangent space}&\map \hbox{tangent space}\cr
\end{eq}E.g. a Lie group $G$  acts adjointly
$G\x G'\map G'$, $\Ad g(l)=glg^{-1}$, upon its Lie algebra
$G'=\log G$ or on its tensor fields.

With the tangent distributions  dual
to the  symmetric space functions the residual product (convolution)
 of a  tangent space  function with a group function
arises in the dual product
\begin{eq}{rl}
\cl G'\x\cl G\map\C,~~
\dprod{ l}{ g}&=\int  l(x)dx ~  g(x)\cr
&=\int d^dx\int d^dq~e^{iqx}(\tilde l*\tilde g)(q)\cr
&=(2\pi)^d(\tilde l*\tilde g)(0)
\end{eq}With $\dprod{ l}{ g}=1$ the tangent and symmetric space functions
are called dual to each other.

\subsection
{Eigenvalue Equations}

The tangent action defines
eigenfunctions for an invariant $\mu\in\C$ by, e.g.
\begin{eq}{l}
{1\over \mu}{d\over dx} g(x)= g(x),~~
 {1\over \mu^2}  {\p^2\over  \p\rvec x^2} g(\rvec x)= g(\rvec x)
,~~ {1\over \mu}  {\p\over  \p\rvec x} g(\rvec x)= g(\rvec x)

\end{eq}The
 invariant is the solution of the {\it eigenvalue equation}
for the massless tangent function $ l_0$, $\de'* l_0=\de$,
(inverse derivative), e.g.
$\tilde  l_0(q)={\mu\over q},{\mu\rvec q\over \rvec q^2},
{\mu^2\over \rvec q^2},{mq\over q^2}$,
with the unit or
the Lorentz compatibly embedded unit  on the r.h.s. - with the examples
\begin{eq}{rrl}
\hbox{time $\R$:}&{m\over q}=1&\then q=m\cr
\hbox{position $\R^3$:}&{\mu\rvec q\over \rvec q^2}=\bl 1_2,~~
{\mu^2\over \rvec q^2}=1
&\then \rvec q^2=\mu^2\cr
\hbox{spacetime $\R^{1+s}$:}&{mq\over  q^2}=\bl1_2,~~
{m^2\over q^2}(\bl 1_{1+s}-{q\ox\d q\over q^2})
=(\bl 1_{1+s}-{q\ox\d q\over q^2})
&\then q^2=m^2\cr
\end{eq}

To obtain invariants for a product representation
a function for a symmetric space representation
acts by  residual product upon the
massless tangent function
leading to another tangent function
 \begin{eq}{l}
\tilde  l_0:\tilde{\cl G}\map \tilde{\cl G}',~~\tilde g\mape
\tilde l_0*\tilde g
 \end{eq}with the
invariant arising from the eigenvalue equation
\begin{eq}{l}
\tilde l_0*\tilde g(q)=1\then q=\mu\cr
\end{eq}This amounts to a normalization of the
$q$-dependent residue arising in a convolution.

E.g., the residual action
of the tangent function ${m\over q}$
of abelian time $\D(1)$
on an irreducible representation ${1\over q-M}$
gives $m+M$ as eigenvalue for the product representation
\begin{eq}{rrl}
{m\over q}:~\R_\od\to \R:&{1\over q-M}\mape {m\over q}~
\stackrel{\rm R}*~{1\over q-M}
&={m\over q-M}\cr
& {m\over q-M}=1&\then q=M+m
\end{eq}

\subsection{Eigenvalues for Position Bound Waves}

The Hamiltonian  for the nonrelativistic hydrogen atom
involves the Kepler
potential  which is a tangent  distribution
arising by Fourier transformation of a massless
function representing  the position translations $\R^3$
\begin{eq}{l}
H={\rvec p^2\over 2}-{1\over r_\ro P},~~
{1\over r_\ro P}=\int {d^3 q\over 2\pi^2}~{1\over \rvec q^2}
e^{-i\rvec q\rvec x}\cr
\end{eq}The eigenvalue equation involves
the residual product with the wave functions $g$ as position representation matrix
elements
\begin{eq}{l}
H g(\rvec x)=E g(\rvec x)\iff
[{\rvec q^2\over 2}-{i\over \rvec q^2}
\hskip.2mm\stackrel{\rm R}*\hskip.7mm] \tilde g(\rvec q)
=E\tilde g(\rvec q)\hbox{ with }\stackrel{\rm R}*={*\over 2i\pi^2}
\end{eq}The residual product of the massless tangent
 representation function ${i\over\rvec q^2}$
with  the  position
representation functions $\tilde g$
with invariant $\mu\in(|m|,\mp i(|m|\pm io))$
 gives tangent representation functions, e.g. for scalar  representations
\begin{eq}{l}
{i\over \rvec q^2}~\stackrel{\rm R}*~\tilde g(\rvec q)=\tilde l(\rvec q)
\hbox{ with }\left\{\begin{array}{rl}
\tilde g(\rvec q)&\in
\{{\SUM_{n=0}^{N}}{a_{-2-n}\over (\rvec q^2+\mu^2)^{2+n}}\mid a_{-2-n}\in\C\}\cr
\tilde l(\rvec q)&
\in\{{\SUM_{n=1}^{N}}{a_{-1-n}\over (\rvec q^2+\mu^2)^{1+n}}\mid a_{-1-n}\in\C\}
\end{array}\right.\cr
\hbox{e.g. for }
g(x)=\int{d^3q\over 2\pi^2}{2\mu\over (\rvec q^2+\mu^2)^2}e^{-i\rvec q\rvec x}
=e^{-\mu r}\then
{i\over \rvec q^2}~\stackrel{\rm R}*~
{2\mu\over (\rvec q^2+\mu^2)^2}
={1\over \rvec q^2+\mu^2}
\end{eq}Therewith the eigenvalue problem can be solved by
noncompact position representation functions (Hilbert space bound waves),
e.g. by
the irreducible scalar position representation
 for the ground state
$\rstate{1,\rvec 0}\sim e^{-r}$
\begin{eq}{l}
[{\rvec q^2\over 2}
-{i\over \rvec q^2}
~\stackrel{\rm R}*~]{1\over (\rvec q^2+\mu^2)^2}
=E {1\over (\rvec q^2+\mu^2)^2}\iff
{\rvec q^2\over 2} -{\rvec q^2+\mu^2\over 2\mu}=E

\then \left\{\begin{array}{rl}
E&=-{\mu^2\over 2}\cr
\mu&=1\cr
\end{array}\right.
\end{eq}Nontrivial knots $N=1,2,\dots$ lead
to  the Laguerre polynomials as linear combinations
of position representation functions.
Analogously, harmonic polynomials
for  angular momenta $L=1,2,\dots$ can be included.

\subsection{The Invariant Mass Ratio for Spacetime}

In general
and in contrast to residual product stable
energy and momentum pole functions,
$\cl P(\ol\C)\stackrel{\rm R}*\cl P(\ol\C)\to\cl P(\ol\C)$,
the residual products of
energy-momentum  $q^2$-pole functions
 for   representations
of rank 2 nonlinear spacetime $\R_\od^{1+s}$
with  hyperbolic-spherical singularity surfaces $q^2=m^2$
do not produce rational complex functions
with $q^2$-poles which would determine
the invariants of product representations.
The $q^2$-dependent residue of the convolution gives
integrals over rational functions, e.g.
\begin{eq}{l}
\hbox{spacetime: }
\int_0^1 d\ze~{1\over q^2\ze-m^2}={\log(1-{q^2\over m^2})\over q^2}
\end{eq}In the following an attempt is made to determine
invariant masses and normalizations of ener\-gy-mo\-men\-tum
poles for  the representations of the time translations
$\R$, Lorentz compatibly
embedded
into   spacetime
 translations $\R^{1+s}$.
Perhaps, one can characterize  this  as an attempt to find
a Lorentz compatible solution of the  bound state problem in the potential
$V_3(r)$ as given above in the projection
of the vector representation of nonlinear spacetime $\R^4_\od$ to
the homogeneous position space
$\SO_0(1,3)/\SO(3)\cong\R^3$.
The superposition of Yukawa and exponential potentials
\begin{eq}{rl}
\rvec F(\rvec x)=-{\p\over\p \rvec x} V_3(r),~~
V_3(r)&
=\int_{m_3^2}^{m_0^2}dm^2(m_0^2-m^2)\int{d^3q\over 2\pi^2}{1
\over (\rvec q^2+m^2)^3}e^{-i\rvec q\rvec x}\cr
&={e^{-  |m_0|r}
-e^{-  |m_3|r}\over r}
+{m_0^2-m_3^2\over 2|m_3|}e^{-  |m_3|r}\cr
\end{eq}is the 2-sphere spread
of a noncompact representation of 1-dimensional position
with a  $z$-proportional contribution from the dipole
(nildimension $N=1$)
\begin{eq}{rl}
V_3(r)=-{d\over dr^2}V_1(r),~~
V_1(z)&
=\int_{m_3^2}^{m_0^2}dm^2(m_0^2-m^2)\int{dq\over \pi}{2
\over (\rvec q^2+m^2)^3}e^{-iqz}\cr
&=\int_{m_3^2}^{m_0^2}dm^2(m_0^2-m^2)({d\over dm^2})^2{e^{-|mz|}\over |m|}\cr
&=2{e^{-  |m_0z|}\over |m_0|}
-\left[2
-{m_0^2-m_3^2\over m_3^2}
(1+|m_3z|)\right]{e^{-  |m_3z|}\over |m_3|}

\end{eq}

The  residual product ${q\over q^2} ~{\stackrel{\rm R}*}~\tilde g$
of the massless vector function for a spacetime tangent  representation
 with the  spacetime vector
representation function $\tilde g$,
characterized by two invariants
 gives
 \begin{eq}{rl}
{q\over q^2}:\R_\od^2\to \R^2:&{q\over q^2}
~{\stackrel{\rm R}*}~
\int_{m_3^2}^{m_0^2}dm^2~
{ q\over(q^2-m^2)^2}\cr
&\hskip4mm=-({1\over 2}\bl1_2+q\ox \d q{\p\over\p q^2})
\int_{m_3^2}^{m_0^2}dm^2\int_0^1 d\ze~{1\over q^2\ze-m^2 }
\cr

{q\over q^2}:\R_\od^4\to \R ^4:&
{q\over q^2}~{\stackrel{\rm R}*}~\int_{m_3^2}^{m_0^2}dm^2(m_0^2-m^2)
{2q \over(q^2-m^2)^3}\cr
&\hskip4mm=-({1\over 2}\bl1_4+q\ox \d q {\p\over\p q2})
\int_{m_3^2}^{m_0^2}dm^2(m_0^2-m^2)
\int_0^1 d\ze  { 1-\ze\over q^2\ze-m^2 }\cr

\end{eq}

The massless tangent function
has a hyperbolic singularity surface.
With at least one
nontrivial invariant, the spacetime representation
function  has a hyperbolic-spherical singularity
surface. Therefore, invariants on the hyperbolic
surface are combined with
invariants on hyperbolic and spherical
surfaces. There is no combination of  invariants
which are both on spherical surfaces.

The  invariant $m^2_0\ne0$ for the normalized embedded time representation
${q\over q^2-m_0^2}$  is used as unit
\begin{eq}{l}
m_0^2\cong 1,~~
{q\over |m_0|}\cong q,~~{m_3^2\over m_0^2}\cong m_3^2
\end{eq}The eigenvalue functions are the $q^2$-dependent
residues
\begin{eq}{l}
\tilde l_\bl 1(q^2)
={M^2({1\over q^2})\over q^2}=\left\{\begin{array}{lll}
-\int_{m_3^2}^{1}{dm^2\over2}&\int_0^1 d\ze~{1\over q^2\ze-m^2}
&\hbox{for }\R^2\cr
-\int_{m_3^2}^{1}{dm^2(1-m^2)\over 2}&
\int_0^1 d\ze  { 1-\ze\over q^2\ze-m^2 }
&\hbox{for }\R^4\cr\end{array}\right.\cr
\cr
\end{eq}

The residual product will be used to
establish duality
between spacetime and   tangent  representation
in the normalization
$({q\over q^2} ~{\stackrel{\rm R}*}~\tilde g)(0)=\bl 1_{1+s}$.
This duality condition requires an eigenvalue at mass
$q^2=0$, i.e. ${\rvec q^2\over q_0^2}=1$,
and determines the ratio
${m_3^2\over m_0^2}$ of the invariants for
rank 2 spacetime
\begin{eq}{l}
1=\tilde l_\bl 1(0)=\left\{\begin{array}{cllll}
-{\log  m_3^2\over2}&\then {m_3^2\over m_0^2}&=e^{-2}&\sim {1\over 7.4}
&\hbox{for }\R^2\cr
-{\log  m_3^2+1-m_3^2\over 4}
&\then {m_3^2\over m_0^2}&\sim e^{-5}&\sim{1\over 148.4}
&\hbox{for }\R^4\cr\end{array}\right.
\end{eq}

\section
[Residues of Tangent Representations]
{Residues of Tangent Representations}

\subsection
[Geometric Transformation and Mittag-Leffler Sum]
{Geometric Transformation\\and Mittag-Leffler Sum}

The exponential  from the Lie algebra $\R$ (time translations) to
the group $\exp \R=\D(1)$
can be reformulated in the  language
of residual representations with energy functions
by a {\it geometric series}
\begin{eq}{rl}
 e^{ im t}
&=\oint {dq \over 2i\pi }{1\over  q-m } e^{i qt}\cr
={\SUM_{k=0}^\infty}{(im t)^k\over k!}
&=\oint {dq\over 2i\pi }{1\over q}
{\SUM_{k=0}^\infty}{m^k\over q^k} e^{i qt}
\end{eq}

The transformations involved
 \begin{eq}{l}
z\mape {1\over z}=w\mape  {w\over 1-w}
={1\over z-1},\hbox{ e.g. }z={q\over m}, {q^2\over m^2}
\end{eq}are  elements  of the broken rational (conformal)
 bijective transformations of the closed complex plane
\begin{eq}{l}
\ol\C\ni z\mape {\al z+\be\over \ga z+\de}\in\ol\C
\end{eq}with real
 coefficients  as group isomorphic to
\begin{eq}{l}
A={\scriptsize \pmatrix{\al&\be\cr\ga&\de\cr}}\in
 \SL(\R^2)\sim \SU(1,1)\sim \SO(1,2)
 \end{eq}For $\det A=1$
upper and lower half plane $x\pm io$ remain stable.
The eigenvalue $w=z=1$ becomes a pole
\begin{eq}{l}
{\scriptsize \pmatrix{\al&\be\cr\ga&\de\cr}}
={\scriptsize \pmatrix{1&0\cr -1&1\cr}} :~~
w\mape {w\over -w+1},~~1\mape\infty,~~0\lrmap 0
\end{eq}With one fixpoint $w=0$
 the transformation is parabolic, i.e.
an element of the $\R$-isomorphic subgroup
${\scriptsize \pmatrix{1&0\cr\ga&1\cr}}$.

The geometric transformation will be generalized to
associate  pole functions
to the complex eigenvalue functions  for spacetime
with $z={q^2\over m^2}$
\begin{eq}{l}
z\mape l(z)\mape {l(z)\over 1-l(z)}
\end{eq}An eigenvalue, i.e. a zero  of the denumerator
$z_0\in\{z\mid l(z)=1\}$ - assumed to be simple with $l$ holomorphic there -
defines, by geometric transformation of its
Taylor series, a {\it Laurent series}\cite{BESO} and a
 residue
 \begin{eq}{rcl}
l(z)=&1+(z-z_0)l'(z_0)&+{\SUM_{k=2}^\infty}
{(z-z_0)^k\over k!}f^{(k)}(z_0)\cr
{l(z)\over 1-l(z)}=&
{a_{-1}(z_0)\over z-z_0}&+{\SUM_{k=0}^\infty}
(z-z_0)^ka_k(z_0)\cr
a_{-1}(z_0)=&-{1\over l'(z_0)}\hfill&
\end{eq}Each eigenvalue $\{z_k\mid l(z_k)=1\}$ has its own
principal part with the {\it Mittag-Leffler sum}
replacing the simple pole for time or position
\begin{eq}{l}
z\mape\l(z)\mape {l(z)\over 1-l(z)}\mape {\SUM_{z_k}}{\al_{-1}(z_k)\over z-z_k}
\end{eq}The generalization for  higher order poles is obvious.

Therewith one obtains
the transition from the eigenvalue function
$\tilde l_0*\tilde g$
to    complex  representation functions
for  the Poincar\'e group
\begin{eq}{l}
{\tilde l_0*\over 1-\tilde l_0*}:\tilde{\cl G}\map \tilde{\cl G}',~~~\left\{\begin{array}{rl}
\tilde g&\mape\tilde l_0*\tilde g\cr
\tilde l_0*\tilde g(q^2)&=1+(q^2-m^2){d\over dq^2}\tilde l_0*\tilde g(m^2)+\dots\cr
 \tilde g(q^2)&\mape {\tilde l_0*\tilde g(q^2)\over 1- \tilde l_0*\tilde g(q^2) }
={a_{-1}(m^2)\over q^2-m^2}+\dots\cr
-{1\over a_{-1}(m^2)}&={d\over dq^2}\tilde l_0*\tilde g(m^2)
\end{array}\right.
\end{eq}

\subsection{Residues as Coupling Constants}

For the residual spacetime product above
$({q\over q^2}~{\stackrel{\rm R}*}~\tilde g)(q^2)$
the residual normalization $a_{-1}(0)$
for the massless solution
$\tilde l_\bl 1(0)=1$ is given
by the inverse of the  negative derivative of the
eigenvalue function there
\begin{eq}{l}
-{1\over a_{-1}(0)}=
{\p\over\p q^2}\tilde l_\bl 1(0)=\left\{\begin{array}{clll}
{1-m_3^2\over 4m_3^2}&={e^2-1\over 4}&\sim 1.6&\hbox{for }\R^2\cr
{1-6m_3^2+m_3^4\over 12 m_3^2}&\sim{e^5\over 12}&\sim 12.4
&\hbox{for }\R^4\cr\end{array}\right.
\end{eq}

With the  geometric  transformation the principal part in
the Laurent series  gives
 an ener\-gy-mo\-men\-tum spacetime translation representation function
for  mass zero
with residual normalization
 \begin{eq}{rll}
{{q\over q^2}*\over 1-{q\over q^2}*}:\R^2_\od\map\SO_0(1,1)\sx\R^2:&\int_{m_3^2}^1dm^2~
{ q\over(q^2-m^2)^2}
&\mape \bl 1_2 {a_{-1}(0)\over q^2}+\dots\cr
{{q\over q^2}*\over 1-{q\over q^2}*}:\R^4_\od\map\SO_0(1,3)\sx\R^4:&\int_{m_3^2}^1dm^2~
{ (1-m^2)2q\over(q^2-m^2)^3}
&\mape \bl 1_4 {a_{-1}(0)\over q^2}+\dots
\end{eq}With appropriate integration contour, it can be  used as
propagator for a mass zero spacetime vector field
 with coupling constant $-a_{-1}(0)$ which -
 with the signature $s-(d-s)$ only for 4-dimensional spacetime -
 has two particle interpretable
  degrees of freedom
  with a positive scalar product, related to the 2-sphere
   $\R^4_\od/\R^2_\od\cong\Om^2$ with left and right axial $\SO(2)$-rotations
   (polarization)
 \begin{eq}{l}
 -\eta^{jk}=\left\{\begin{array}{rll}
 {\scriptsize\pmatrix{-1&0\cr0&1\cr}}&\cong {\scriptsize\pmatrix{0&1\cr 1&0\cr}}
 &\hbox{for }\SO_0(1,1)\sx\R^2\cr
 {\scriptsize\pmatrix{-1&0\cr0&\bl 1_3\cr}}&
 \cong {\scriptsize\pmatrix{0&0&1\cr
 0&\bl1_2&0\cr1&0&0\cr}}
 &\hbox{for }\SO_0(1,3)\sx\R^4\cr
\end{array}\right.
\end{eq}

All the numerical results depend on the normalizations -
trace normalization, dual normalization - which
require a deeper understanding. If those normalizations can be trusted
and if appropriate representations of the compact
internal  degrees of freedom
for $\U(2)$ hypercharge and isospin are included,
the residue of the arising  propagator with mass zero in 4-dimensional spacetime
 may be compared with
the coupling constant\cite{HEI}
 in the propagator of a massless gauge field, e.g. for the electromagnetic
interaction and the
left and right polarized photons with Sommerfeld's fine structure constant $\al$
\begin{eq}{l}
\SO_0(1,3)\sx\R^4:~~ -\eta^{jk}{e^2(0)\over q^2}
\hbox{ with }{1\over e^2(0)}={1\over 4\pi\al}\sim{137\over 12.6}\sim 10.9
\end{eq}

\end{document}